\documentclass[apj]{emulateapj}
\usepackage{multirow}
\usepackage{booktabs}
\usepackage{graphicx}
\usepackage{amsmath}
\usepackage{longtable}
\usepackage{rotating}
\usepackage{enumerate}

\newcommand{\msun}{M_\odot}

\newcommand{\me}{M_\oplus}

\newcommand{\mth}{M_{\rm th}}

\newcommand{\rp}{r_{\rm p}}
\newcommand{\lsh}{l_{\rm sh}}
\newcommand{\rgap}{r_{\rm gap}}
\newcommand{\rring}{r_{\rm ring}}
\newcommand{\mplanet}{M_{\rm p}}

\newcommand{\cs}{c_{\rm s}}
\newcommand{\imm}{I_{\rm mm}}
\newcommand{\ts}{\tau_{\rm stop}}
\newcommand{\tmidplane}{T_{\rm midplane}}
\newcommand{\sigmag}{\Sigma_{\rm gas}}
\newcommand{\sigmad}{\Sigma_{\rm dust}}

\begin{document}
\title{Multiple Disk Gaps and Rings Generated by a Single Super-Earth:\\ II. Spacings, Depths, and Number of Gaps, with Application to Real Systems}

\author{Ruobing Dong\altaffilmark{1,2}, Shengtai Li\altaffilmark{3}, Eugene Chiang\altaffilmark{4}, \& Hui Li\altaffilmark{3}}

\altaffiltext{1}{Steward Observatory, University of Arizona, Tucson, AZ, 85719; rbdong@gmail.com}
\altaffiltext{2}{Department of Physics \& Astronomy, University of Victoria, Victoria, BC, V8P 1A1, Canada}
\altaffiltext{3}{Theoretical Division, Los Alamos National Laboratory, Los Alamos, NM 87545}
\altaffiltext{4}{Department of Astronomy, University of California at Berkeley, Berkeley, CA 94720}

\begin{abstract}
ALMA has found multiple dust gaps and rings in a number of protoplanetary disks in continuum emission at millimeter wavelengths. The origin of such structures is in debate. Recently, we documented how one super-Earth planet can open multiple (up to five) dust gaps in a disk with low viscosity ($\alpha\lesssim10^{-4}$). In this paper, we examine how the positions, depths, and total number of gaps opened by one planet depend on input parameters, and apply our results to real systems. Gap locations (equivalently, spacings) are the easiest metric to use when making comparisons between theory and observations, as positions can be robustly measured. We fit the locations of gaps empirically as functions of planet mass and disk aspect ratio. We find that 
the locations of the double gaps in HL Tau and TW Hya, and of
all three gaps in HD 163296, are consistent with being opened by a sub-Saturn mass planet. This scenario predicts the locations of other
gaps in HL Tau and TW Hya,
some of which appear consistent with current observations.
We also show how the Rossby wave instability may develop at the edges of several gaps and result in multiple dusty vortices, all
caused by one planet. A planet as low in mass as Mars may produce multiple dust gaps in the terrestrial planet forming region. 
\end{abstract}



\section{Introduction}\label{sec:intro}

The Atacama Large Millimeter Array (ALMA) has discovered multiple gaps and rings in a number of protoplanetary disks in dust continuum emission at 
millimeter (mm) wavelengths at $\sim$10--100 AU. Examples include the disk around HL Tau \citep{brogan15}, TW Hya \citep{andrews16, tsukagoshi16}, HD 163296 \citep{isella16hd163296}, AS 209 \citep{fedele18}, AA Tau \citep{loomis17}, Elias 24 \citep{cieza17, cox17, dipierro18}, GY 91 \citep{sheehan18}, and V1094 Sco \citep{ansdell18, vanterwisga18}. The origins of these gaps are being debated. Several hypotheses have been put forward, including disk-planet interaction \citep[e.g.,][]{dong15gap, dipierro15hltau, pinilla15twoplanets, jin16}, zonal flows \citep{pinilla12dusttrapping}, secular gravitational instability \citep[e.g.,][]{takahashi14}, 
self-induced
dust pile-ups \citep{gonzalez15}, radially variable magnetic disk winds \citep{suriano17, suriano18}, and dust sintering and evolution at snowlines \citep{zhang15, okuzumi16}. In nearly all mechanisms, perturbations in gas are followed 
by radial drift and concentration of 
mm-sized dust \citep[e.g.,][]{weidenschilling77, birnstiel10, zhu12}. 
In planet-disk interaction scenarios,
even super-Earth planets 
may 
open dust gaps \citep[e.g.,][]{zhu14votices,rosotti16, dipierro16, dipierro17}.

At first glance, multiple gaps seem to imply multiple planets,
with one planet embedded within each gap. Recently, however, 
it has been appreciated that a single planet can generate multiple gaps. Using gas+dust simulations, \citet{dong17doublegap} showed that one super-Earth planet 
(with mass between Earth and Neptune) 
can produce up to five dust gaps in a nearly inviscid disk, and that these gaps would be
detectable by ALMA  (see also \citealt{muto10, duffell12, zhu13, bae17, chen18, ricci18}). 
In particular, a pair of closely spaced narrow gaps (a ``double gap'') sandwiches the planet's orbit. This structure is created by the launching, shocking, and dissipation of two primary density waves, 
as shown in \citet[see also \citealt{rafikov02, rafikov02migration}]{goodman01}. 
Additional gaps interior to the planet's orbit may also be present, possibly opened
by the dissipation of secondary and tertiary density waves
excited by the planet \citep{bae18theory,bae18simulation}.

Motivated
by the 
boom of multi-gap structures discovered by ALMA,
and the mounting evidence of extremely low levels of turbulence at the disk midplane (from direct gas line observations, e.g., \citealt{flaherty15, flaherty17, teague18twhya}, and from evidence for dust settling, e.g.,  \citealt{pinte16,stephens17}), we 
investigate here what properties of the planet and disk can be inferred
from real-life gap observations. 
We study systematically how gap properties depend on planet and disk parameters, 
and propose generic guidelines connecting ALMA observations with disk-planet interaction models. We introduce our models in \S\ref{sec:simulation}, and present our main parameter survey of gap properties in \S\ref{sec:results}. These results are applied to three actual disks in \S\ref{sec:applications}, 
discussed
in \S\ref{sec:discussions}, and summarized in \S\ref{sec:summary}.


\section{Simulations}\label{sec:simulation}

We use two-fluid (gas + dust) 2D hydrodynamics simulations 
in cylindrical coordinates (radius $r$ and azimuth $\phi$)
to follow the evolution of gas and dust in a disk containing
one planet. The code used is 
\texttt{LA-COMPASS} (\citealt{li05, li09}). 
The numerical setup, largely adopted from \citet{dong17doublegap}, is summarized here (more details can be found in \citealt{fu14} and \citealt{miranda17}).

We adopt a locally isothermal equation of state: pressure $P=\sigmag\cs^2$, where $\sigmag$ and $\cs$ are the surface density and sound speed of the gas. Dust is modeled as a pressureless second fluid. The following Navier-Stokes equations are solved:
\begin{align}
\frac{\partial \sigmag}{\partial t} + \nabla \cdot (\sigmag \mathbf{v}_{\rm gas}) =  0, \\
\frac{\partial \sigmad}{\partial t} +\nabla \cdot (\sigmad \mathbf{v}_{\rm dust}) = \nonumber \\
\nabla \cdot \left[\sigmag D_{\rm dust} \nabla \left(\frac{\sigmad}{\sigmag}\right)\right],  \\
\label{eq:gasmomentum}
\frac{\partial_t \sigmag\mathbf{v}_{\rm gas}}{\partial t} +\nabla(\mathbf{v}_{\rm gas}\cdot\sigmag \mathbf{v}_{\rm gas})+\nabla P = \nonumber \\
-\sigmag\nabla\Phi_{\rm G}+\sigmag \mathbf{f}_{\nu}-\sigmad \mathbf{f}_{\rm drag},  \\ 
\frac{\partial \sigmad\mathbf{v}_{\rm dust}}{\partial t} +\nabla(\mathbf{v}_{\rm dust}\cdot\sigmad \mathbf{v}_{\rm dust}) =\nonumber \\
 -\sigmad\nabla\Phi_{\rm G}+\sigmad \mathbf{f}_{\rm drag},
\end{align}
where $D_{\rm dust}$ is the dust diffusivity defined as $D_{\rm dust}=\nu/(1+\ts^2)$ \citep{takeuchi02}, $\Phi_{\rm G}$ is the gravitational potential 
(including both the gravity of the planet and the self-gravity of the disk; a smoothing length of 0.7$h$ is applied to the gravitational potential of the planet), $\mathbf{f}_{\rm drag}=\frac{\Omega_{\rm K}}{\ts}(\mathbf{v}_{\rm gas}-\mathbf{v}_{\rm dust})$ is the drag force on dust by gas, $\ts$ is the dimensionless stopping time (see below), and $\mathbf{f}_{\nu}$ is the Shakura-Sunyaev viscous force with $\nu=\alpha h^2\Omega_{\rm K}$, $h$ the disk scale height, $\Omega_{\rm K}$ the Keplerian angular frequency, and $\alpha < 1$ a dimensionless constant. Other symbols have their usual meanings. The last term in Eqn.~(\ref{eq:gasmomentum}) accounts for the dust back-reaction on gas. 

The initial surface densities of gas and dust are (we use the subscript 0 to indicate initial conditions):
\begin{align}
\Sigma_{\rm gas,0}=& 5.55\left(\frac{\rp}{r}\right)^{0.5}\ {\rm g\ cm^{-2}},\\
\Sigma_{\rm dust,0}=&  10^{-2}\Sigma_{\rm gas,0},
\end{align}
where $\rp=30$~AU is the radius of the planet's circular orbit, which is fixed in our simulations. We note that if a planet migrates in the radial direction fast enough, 
gap properties can be affected \citep[Fig. 14]{dong17doublegap}. However, low-mass planets in  near-invicsid environments as studied here are not expected to 
migrate significantly
\citep{li09, yu10, fung17}. We ramp up the planet mass with time $t$ as $\mplanet(t)\propto \sin{(t)}$ over 10 orbits. In \citet[Fig. 6]{dong17doublegap} we experimented with different planet growth timescales up to a few thousand orbits, and found no significant effects on the rings and gaps ultimately produced by the planet.

The disk aspect ratio $h/r$ is assumed radially invariant. We adopt a constant instead of a flared $h/r$ profile to 
suppress 
the dependences of gap properties on the 
$h/r$ gradient, and to increase the timestep at the inner boundary. We note that fractional gap spacings, which we will use as our primary
diagnostic of models and observations (see our equation \ref{eq:spacing}), 
depend only weakly 
on radial gradients in 
$h/r$, $\Sigma_{\rm gas,0}$, and the slope of the background
$\sigmag(r)$ profile. \citet[Fig. 8]{dong17doublegap} showed that as the background $\sigmag(r)$ profile varies from $\sigmag\propto r^0$ to $\sigmag\propto r^{-1.5}$, all gaps and rings shift inward, while their relative separations are unchanged at the level of a few percent. 

The dimensionless stopping time of dust --- the particle momentum stopping time
normalized to the dynamical time --- is, in the Epstein drag regime,
\begin{equation}
\ts=\frac{\pi s \rho_{\rm particle}}{2\sigmag},
\label{eq:stop}
\end{equation}
where $\rho_{\rm particle}=1.2$ g cm$^{-3}$ is the internal density of dust particles, and $s=0.2$~mm is the particle size. Such particles are probed by continuum emission at mm wavelengths \citep[e.g.,][]{kataoka17}. At $r=0.2$--$1 \,\rp$, $\ts$ is on the order of 0.01. We note that for mm-to-cm sized particles often traced by VLA cm continuum observations (possibly having $\ts\sim0.1-1$),  gaps and rings similar to those in 0.2 mm sized dust are generated. Qualitatively, as particle size increases, gaps widen and deepen with their locations largely unchanged; the amplitude of the middle ring (MR) co-orbital with the planet drops; particle concentrations at the triangular Lagrange points L$_4$ and L$_5$ become stronger; and an azimuthal gap centered on the planet's location emerges and widens \citep{lyra09, zhu14votices, rosotti16, ricci18}.

The simulation domain runs from 0 to $2\pi$ in $\phi$, and 0.1 to 2.1$\rp$ in $r$. The initial gas disk mass is $0.015\msun$. We set up a wave damping zone at $r=0.1$--$0.2\rp$ as described in \citet{devalborro06}, and an inflow boundary condition at the outer boundary. The resolution
is discussed in \S\ref{sec:resolution}.

We summarize our models in Table~\ref{tab:models}. The model name indicates the
values of $h/r$ and $\mplanet / \mth$ adopted (e.g., in Model H003MP02, $h/r=0.03$ and $\mplanet=0.2\mth$). 
In all cases the planet mass $\mplanet$ is lower than the disk thermal mass $\mth=M_\star(h/r)^3$ ($M_\star = M_\odot$ is the stellar mass); the size of the planet's Hill sphere is comparable to $h$ when $\mplanet=\mth$. An extremely low viscosity is necessary to enable low-mass planets to open gaps. We adopt $\alpha=5\times10^{-5}$ for models with $\mplanet>0.1\mth$, where $\alpha$ is the Shakura \& Sunyaev parameter characterizing the disk viscosity; in those models with $\mplanet\leq0.1\mth$, we use even lower viscosities $\alpha=10^{-6}$--$10^{-5}$ to speed up gap opening \citep[cf.,][]{fung14}.

\begin{table*}[]
\centering
\begin{tabular}{@{}ccccccc@{}}
\toprule
Models & Purpose & $h/r$ & $\mplanet$$^a$ & Grid & Orbits & \# of Gaps \\ \midrule
H002MP02 & \multirow{7}{*}{$h/r$} & 0.02 & $0.2\mth=0.5\me$ & $9216\times6144$ & 1400 & 5 \\
H003MP02 &  & 0.03 & $0.2\mth=1.8\me$ & $4608\times3072$ & 1100 & 5 \\
H004MP02 &  & 0.04 & $0.2\mth=4.3\me$ & $4608\times3072$ & 850 & 4 \\
H005MP02 &  & 0.05 & $0.2\mth=8.3\me$ & $4608\times3072$ & 800 & 3 \\
H006MP02 &  & 0.06 & $0.2\mth=14\me$ & $3072\times3072$ & 750 & 3 \\
H007MP02 &  & 0.07 & $0.2\mth=23\me$ & $3072\times3072$ & 700 & 3 \\
H008MP02 &  & 0.08 & $0.2\mth=34\me$ & $3072\times1536$ & 700 & 3 \\ \midrule
H003MP005 & \multirow{4}{*}{$\mplanet$}  & 0.03 & $0.05\mth=0.45\me$ & $4608\times3072$ & 10700 & 3 \\
H003MP01 &  & 0.03 & $0.1\mth=0.9\me$ & $4608\times3072$ & 2800 & 5 \\
H003MP04 &  & 0.03 & $0.4\mth=3.6\me$ & $4608\times3072$ & 270 &  5 \\
H003MP08 &  & 0.03 & $0.8\mth=7.2\me$ & $4608\times3072$ & 90 & 5 \\ \midrule
H005MP08 & \multirow{3}{*}{Misc.} & 0.05 & $0.8\mth=33\me$ & $4608\times3072$ & 70 & 4 \\ 
Mars & & 0.02 & $0.04\mth=0.1\me$ & $9216\times6144$ & $>$20,000$^b$ & 4 \\ 
RWI &  & 0.08 & $0.4\mth=68\me$ & $1536\times1536$ & 140 & 3 \\ 
\bottomrule
\end{tabular}
\caption{Model Grid. The second to last column lists the number of orbits it takes for OG1 to reach 50\% depletion in dust. The last column is the number of gaps between $r=0.4\rp$ and OG1 (inclusive). $^a$Assuming $\mplanet=\msun$. $^b$OG1 in Model Mars does not reach 50\% depletion in dust when the simulation is terminated at 20,000 orbits.}
\label{tab:models}
\end{table*}

\begin{figure*}
\begin{center}
\includegraphics[trim=0 0 0 0, clip,width=0.8\textwidth,angle=0]{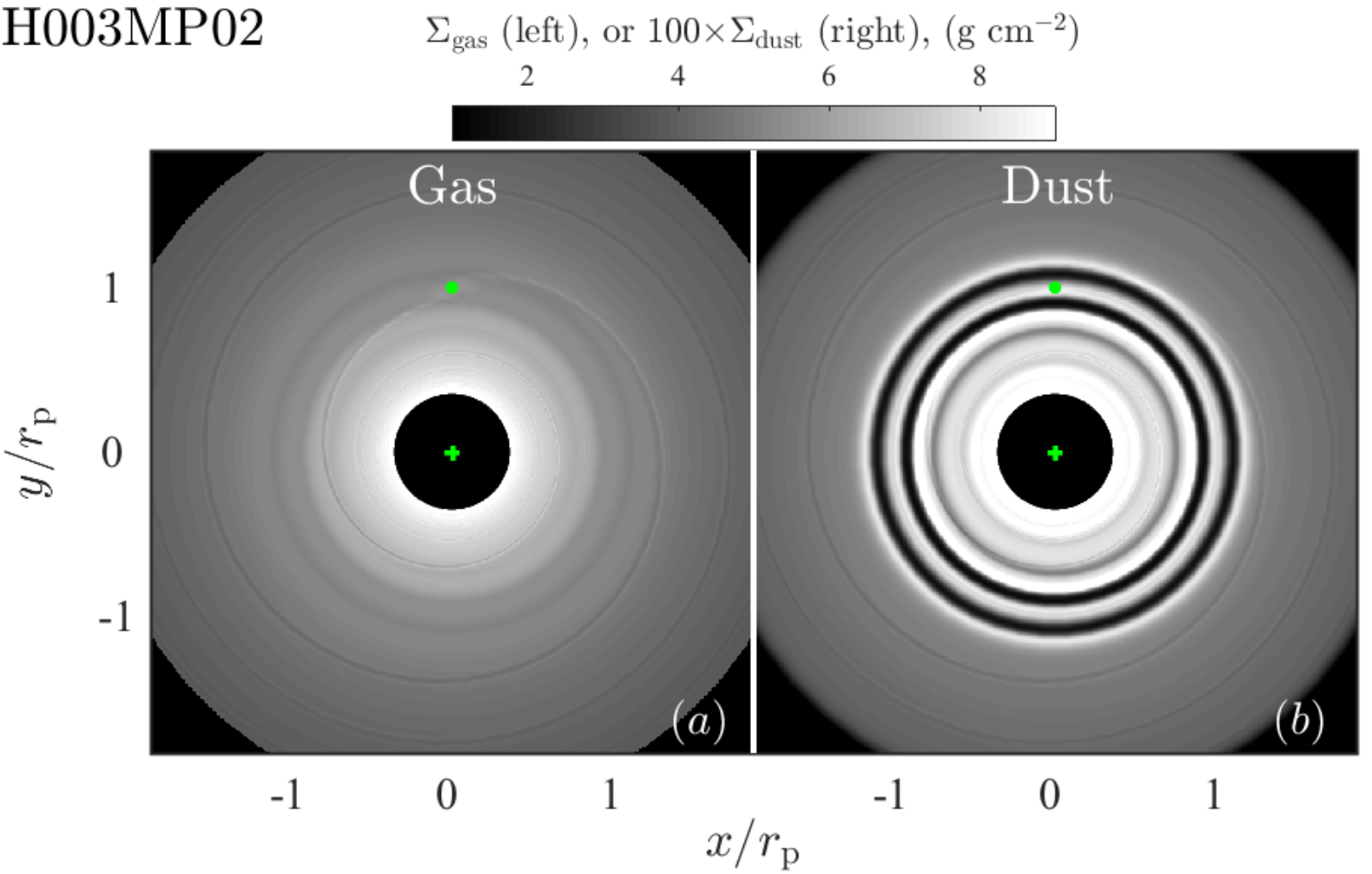}
\includegraphics[trim=0 0 0 0, clip,width=0.5\textwidth,angle=0]{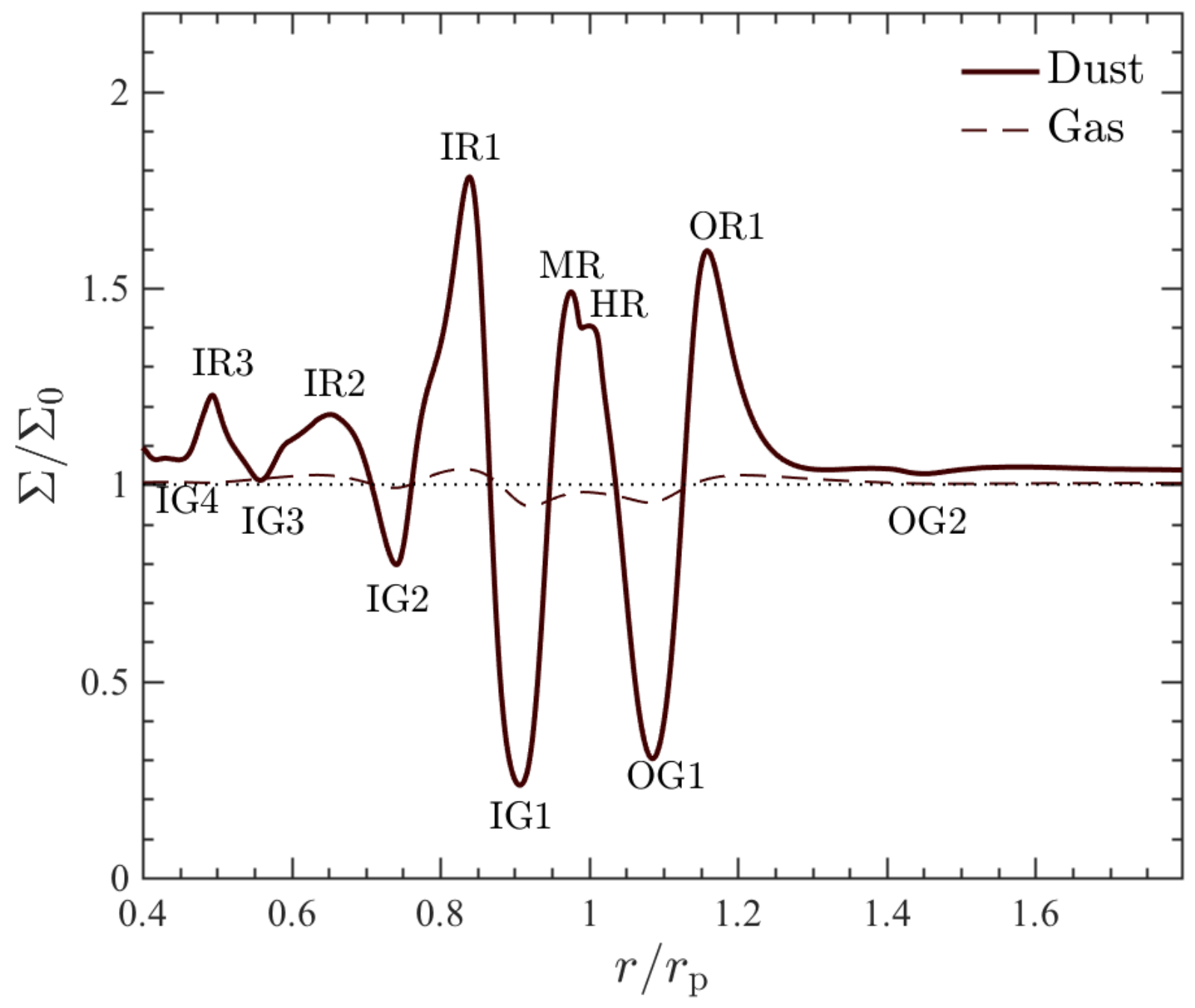}
\end{center}
\figcaption{{\it Top:} Surface density maps of gas and dust in a representative model (H003MP02) at 1600 orbits (0.26 Myr at 30 AU), showing multiple gaps and rings produced by a 1.8$\me$ planet. The green plus and dot mark the locations of the star and the planet, respectively. The inner region is manually masked out. {\it Bottom:} Azimuthally averaged radial profiles of surface density perturbations in dust (solid line) and gas (dashed line) with gaps and rings labeled: IR$n$ (the $n$-th inner ring from the planet), IG$n$ (the $n$-th inner gap), MR (middle ring), HR (the horseshoe ring), ORn (the $n$-th outer ring), and OG$n$ (the $n$-th outer gap). Density fluctuations in gas are much smaller than in dust. See \S\ref{sec:simulation} for details.
\label{fig:example}}
\end{figure*}

Figure~\ref{fig:example} shows the 2D gas and dust distributions and their azimuthally averaged 1D radial profiles for Model H003MP02 at 1600 orbits (0.26 Myr at 30 AU). A $0.2\mth = 1.8 \me$ planet produces multiple gaps and rings in both gas and dust, while the surface density perturbations in 
dust are
an order of magnitude larger than they are in gas.
In each of the dust rings, a few Earth masses of dust are trapped. 
We refer to the gaps and rings inside $\rp$ as ``inner'' gaps and rings (IG and IR), and the ones outside $\rp$ as ``outer'' gaps and rings (OG and OR), and number them according to their distances from the planet. 

\subsection{Numerical Resolution in Hydro Simulations}\label{sec:resolution}

Sufficient spatial resolution is needed to properly capture dust gaps,
as resolution affects numerical diffusivity.
In inviscid or nearly inviscid disks,
the angular momentum carried by a wave launched by a
planet should be conserved 
until the wave shocks and deposits its angular momentum locally, and
opens a gap \citep[see the weakly nonlinear theory of density waves;][]{goodman01}.
Insufficient resolution leads to excessive numerical diffusion; gaps
either appear too close to the planet and are too shallow
\citep{yu10, dong11linear, dong11nonlinear}, or fail to be
seen at all \citep{bae17}.

\begin{figure*}
\begin{center}
\includegraphics[trim=0 0 0 0, clip,width=\textwidth,angle=0]{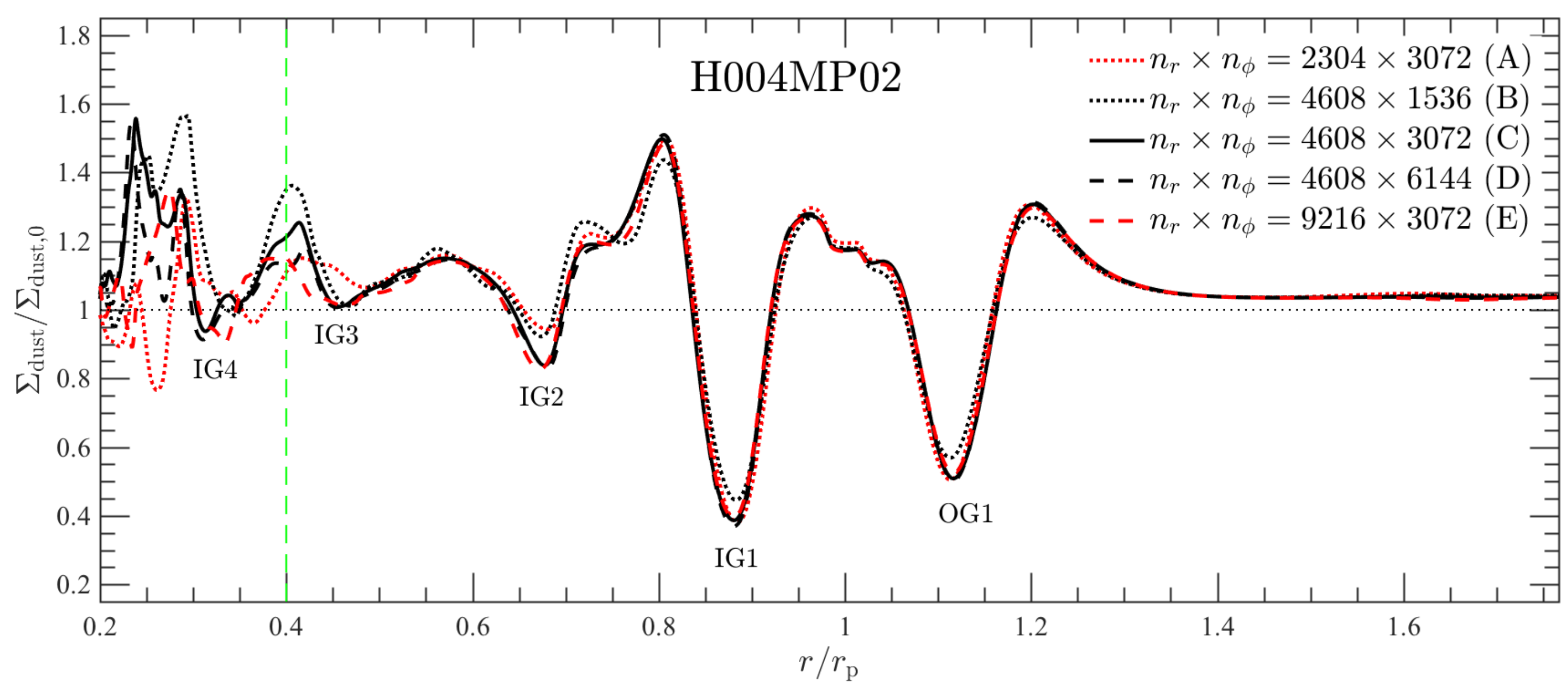}
\end{center}
\figcaption{Azimuthally averaged radial profiles of $\sigmad/\Sigma_{\rm dust,0}$ for five H004MP02 runs at 850 orbits at different resolutions. Runs A, C, and E successively double the number of radial cells ($n_r$), while runs B, C, and D successively double the number of azimuthal cells ($n_\phi$). The profiles at $r\gtrsim0.4\rp$ converge with a grid of $4608\times3072$ or more cells (C, D, E). See \S\ref{sec:resolution} for details.
\label{fig:resolutionrp}}
\end{figure*}

\begin{figure*}
\begin{center}
\includegraphics[trim=0 0 0 0, clip,width=\textwidth,angle=0]{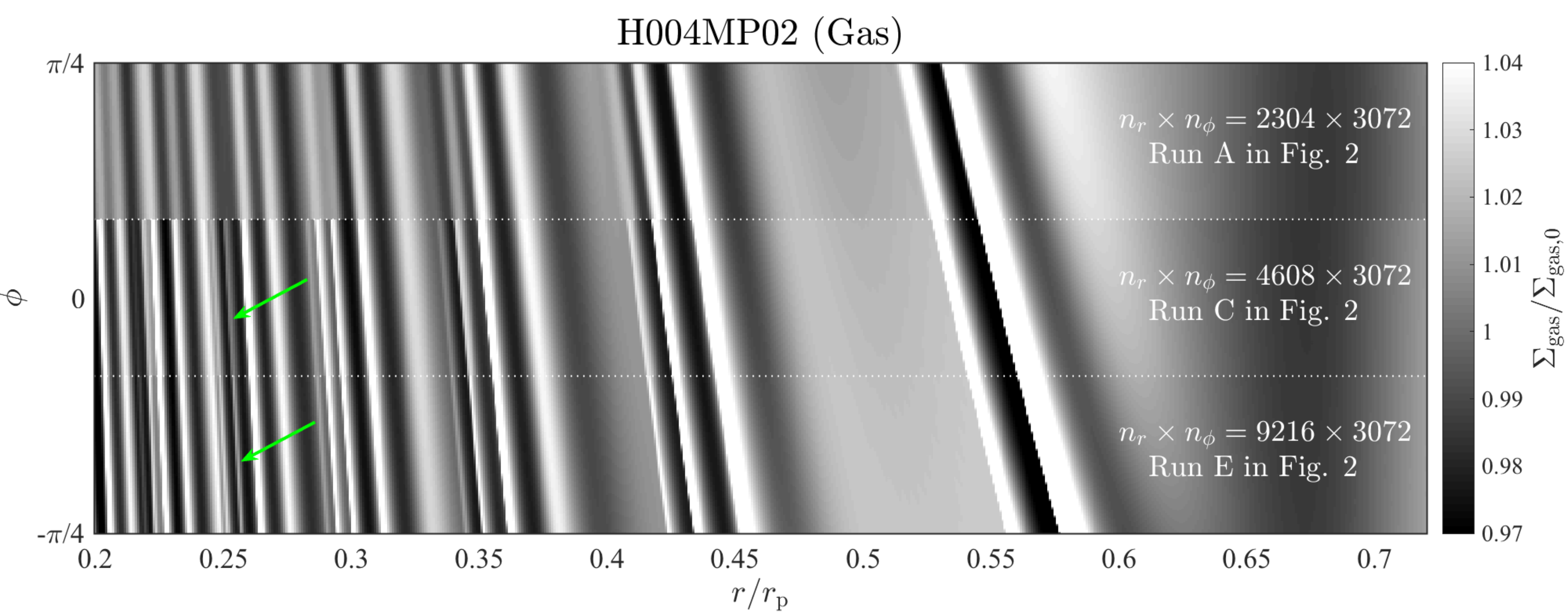}
\includegraphics[trim=0 0 0 0, clip,width=\textwidth,angle=0]{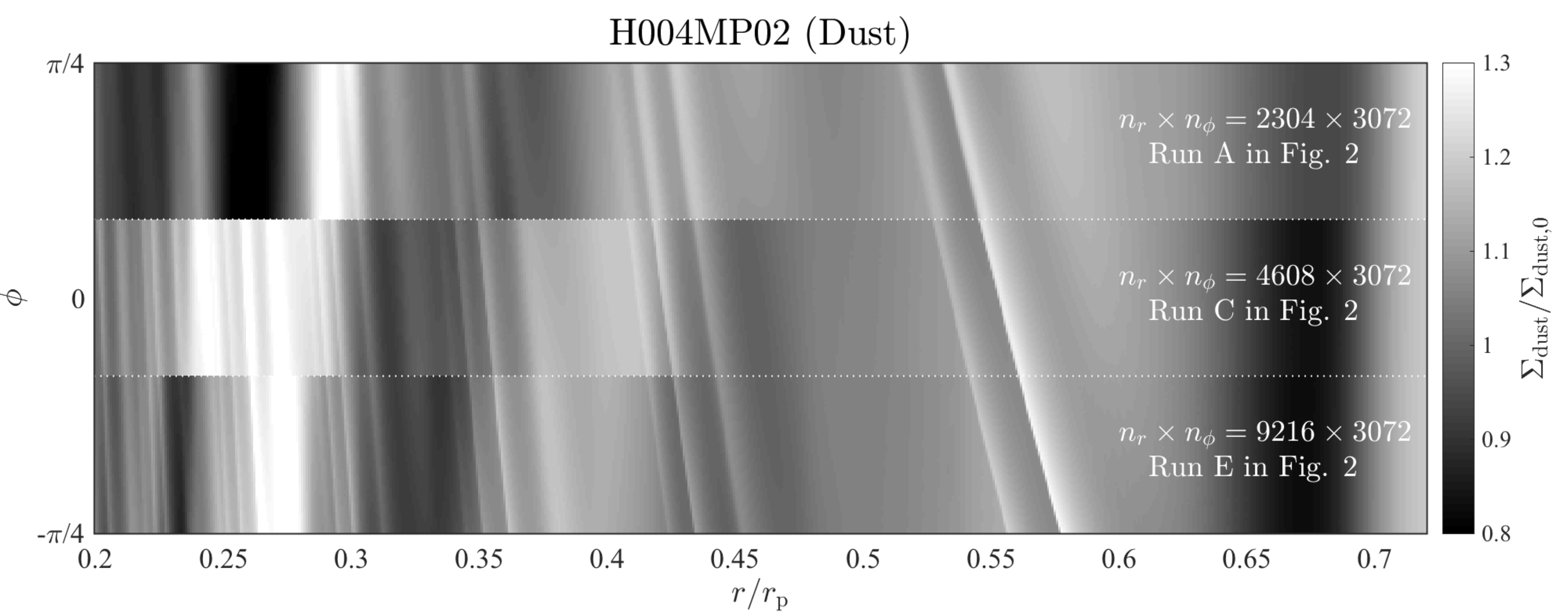}
\end{center}
\figcaption{The dependence of 2D density structures on the resolution in hydro simulations. The  top (bottom) panel shows $\sigmag/\Sigma_{\rm gas,0}$ ($\sigmad/\Sigma_{\rm dust,0}$) maps in polar coordinates for runs A, C, and E in Figure~\ref{fig:resolutionrp}. Each run corresponds to one horizontal stripe, and adjacent runs are separated by a dotted line. Note that each map only covers 1/4 of the simulation domain in the azimuthal direction (1/12 for each run). The planet is located at ($r,\phi$)=($\rp,0$) outside the box to the right. In the dust maps (bottom), annular gaps and rings are vertical stripes. If the adjacent two runs achieve convergence, their wave and gap patterns should smoothly join each other at the boundary. The green arrows in the upper panel mark a density wave present in runs C and E but not in A. Simulations with insufficient spatial resolutions produce a smaller number of density waves in gas, and significantly different dust distributions. See \S\ref{sec:resolution} for details.
\label{fig:resolutionsigma}}
\end{figure*}

Figure~\ref{fig:resolutionrp} compares the radial profiles of
$\sigmad/\Sigma_{\rm dust,0}$ from five runs of Model H004MP02,
labeled A, B, C, D, \& E in order of increasing spatial resolution.
Runs A, C, and E successively double the number of radial cells ($n_r$)
used, while runs B, C, and D successively
double the number of azimuthal cells ($n_\phi$).
The three runs with $n_r\geq4608$ and $n_\phi\geq3072$  (C, D, and E) achieve
visual convergence at $r\gtrsim0.4\rp$. 
Note that insufficient resolutions may also result in a reduced number of 
detectable
density waves \citep[Figure~\ref{fig:resolutionsigma}, see also][]{bae17}.
A higher resolution is needed in the radial direction because density waves
have higher spatial frequencies in the radial direction,
and gaps are radial structures.
At $n_r\times n_\phi=4608\times3072$ for Model H004MP02,
the Hill sphere of the planet is resolved by
75 and 16 cells in the radial and azimuthal directions, while
$h$ is resolved by 96 and 20 cells, respectively.

For all models shown in Table \ref{tab:models},
we adopt resolutions necessary to ensure
visual convergence of results for all disk radii
beyond at least $0.4\rp$ ($2\times$ the outer radius of the
wave damping zone). For runs with $h/r > 0.05$, the resolution
requirements are less severe, and we achieve convergence
for radii outside $0.3\rp$.


\section{Results}\label{sec:results}

In this section, we study how the properties of gaps and rings vary with input parameters. Most simulations are run for $\sim$10$^3$--$10^4$ orbits ($\sim$0.1--1 Myr at 30 AU), after
which time dust gaps are depleted by a factor of $\sim$2--10
in surface density, reminiscent of observed gap depths \citep[e.g.,][of course, observed gap depths are measured in units of surface brightness]{brogan15, andrews16}. Dust gaps in our simulations do not have detectable eccentricities. The locations of gaps and rings, $\rgap$ and $\rring$, are defined as the radii at which $\sigmad/\Sigma_{\rm dust,0}$ attains local minima and maxima, respectively; only robust (non-transient) features are counted. The relative separation (spacing) between two radii (as drawn from $\rgap$, $\rring$, or $\rp$)
is defined as
\begin{equation}
\Delta_{\rm structure1,structure2}=\left|\frac{r_{\rm structure2}-r_{\rm structure1}}{(r_{\rm structure2}+r_{\rm structure1})/2}\right|.
\label{eq:spacing}
\end{equation}
In real observations, $\rp$ is usually unknown, and the spacings of gaps and rings are the direct observables.

\subsection{General Comments on Model-Observation Comparisons}\label{sec:comments}

There are at least three observable gap properties: gap
depth (contrast with adjacent rings), gap width, and gap location.
In model-observation comparisons, gap depth and width are
not easy metrics to work with.
First, narrow gaps are often unresolved in
observations, sabotaging measurements of true depth and width
--- in the \cite{brogan15} observations of HL Tau, the observed
gap widths are comparable to the beam size
\citep[Table~2]{akiyama16hltau}. Second,
spectral index analyses have shown that bright rings between
gaps may be optically thick
\citep[e.g.,][]{brogan15, liu17, huang18},
which implies that dust surface density
does not translate into a unique surface brightness.
Third, gap spacings are less sensitive to the background $\sigmag$ profile, often hard to determine, than gap depths \citep[Fig. 3]{dong17doublegap}.
Finally, gap depth and width evolve with time (\S\ref{sec:time}),
and how long a (hypothetical) planetary system has
perturbed a real disk is usually unknown.

In contrast, it is simpler and
more robust to use gap locations (spacings) 
when comparing models with observations.
Gap locations evolve less with time (\S\ref{sec:time}), and can be determined relatively accurately in observations even when gaps are not well resolved and adjacent rings are optically thick. The total number of gaps may also provide a diagnostic. \citet[Fig. 3]{bae18simulation} showed that
more density waves can be discerned in colder disks, or in disks
perturbed by lower mass planets.

\subsection{The Time Evolution of $\rgap$ and $\rring$}\label{sec:time}

\begin{figure}
\begin{center}
\includegraphics[trim=0 0 0 0, clip,width=0.475\textwidth,angle=0]{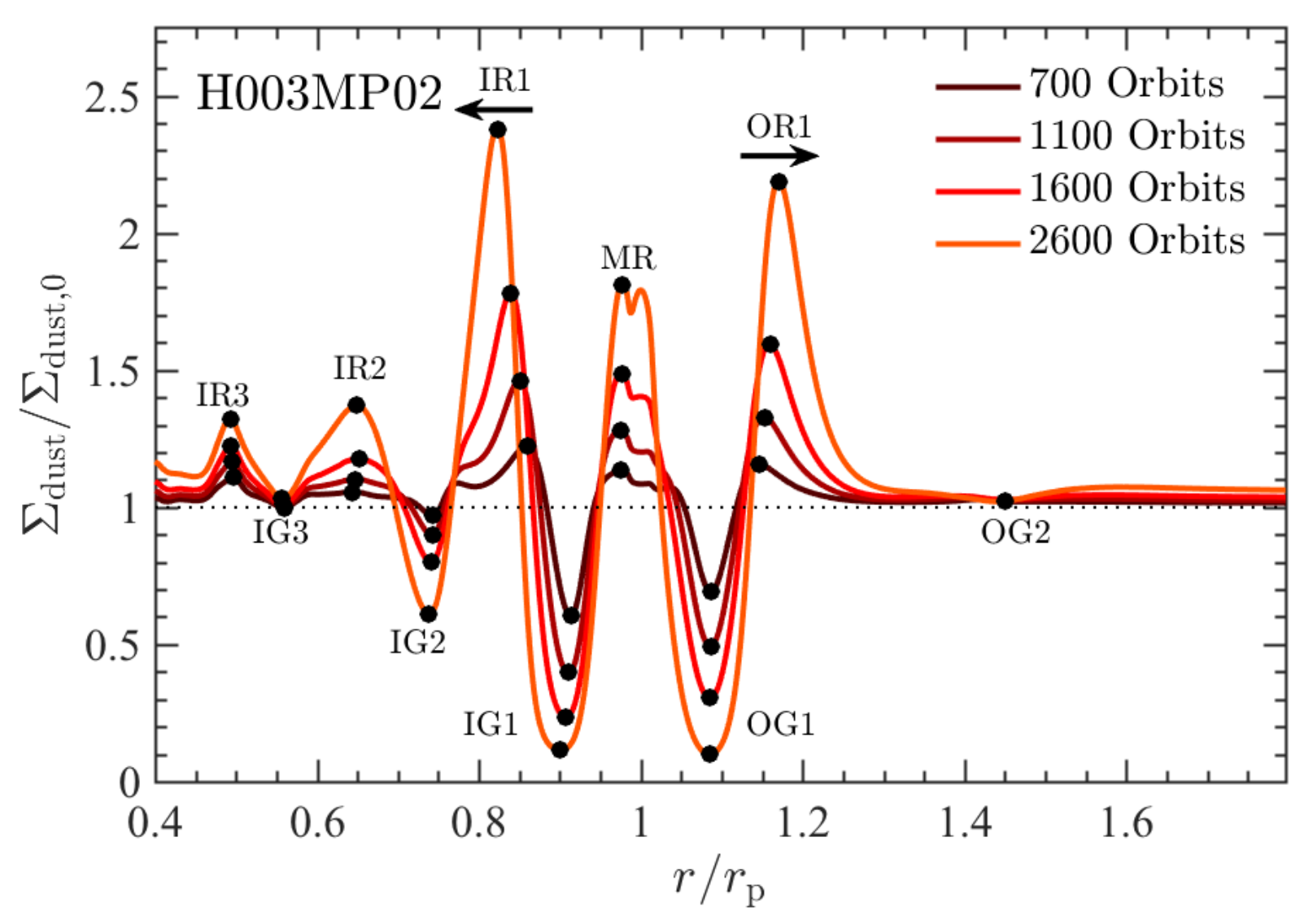}
\end{center}
\figcaption{The radial profile of $\sigmad/\Sigma_{\rm dust,0}$ in Model H003MP02 at four epochs. Local maxima and minima on each profile are marked by dots. $\sigmad$ at OG1 is depleted by 30\%, 50\%, 70\%, and 90\% at 700, 1100, 1600, and 2600 orbits, respectively. The locations of all gaps stay roughly the same, while IR1 and OR1 shift away from the planet, as indicated by arrows. See \S\ref{sec:time} for details.
\label{fig:time}}
\end{figure}

Figure~\ref{fig:time} shows the time evolution of a representative
Model H003MP02. The radial profiles of $\sigmad/\Sigma_{\rm dust,0}$
at 700, 1100, 1600, 2600 orbits are plotted. At these times,
$\sigmad$ at OG2 (outer gap 2) is depleted by 30\%, 50\%, 70\%, and
90\%, respectively. Gap depths are still evolving at 2600 orbits
(0.43 Myr at 30 AU). Rings become more optically thick
and gaps widen with time, making gap contrast and width
problematic metrics in model-observation comparisons.

While gaps deepen noticeably with time, their
locations do not change as much. From 700 to 2600 orbits, IG1
(inner gap 1) moves away from the planet by 15\%; all other gap
locations drift by less than 2\%. In comparison,
the two most prominent dust rings, OR1 (outer ring 1) and IR1 (inner
ring 1), move away from the planet by 17\% and 25\%, respectively.
The fractional separation (a direct observable)
$\Delta_{\rm OG1,IG1}$ increases by 6\%,
while $\Delta_{\rm OR1,IR1}$ increases by 21\% over the course
of the simulation.

We conclude that gap locations (spacings) serve as the most robust
metrics for model-observation comparisons. 
Tables \ref{tab:locations} and \ref{tab:contrasts}
in the Appendix list the locations and values
of $\sigmad/\Sigma_{\rm dust,0}$, respectively, of rings
and gaps in selected models at a time when OG1 is $\sim$50\% 
depleted (for the exact times, see Table \ref{tab:models},
second to last column).

\subsection{The Dependence of Gap Spacing on $h/r$ and $\mplanet$}\label{sec:spacing}

\begin{figure*}
\begin{center}
\includegraphics[trim=0 0 0 0, clip,width=0.7\textwidth,angle=0]{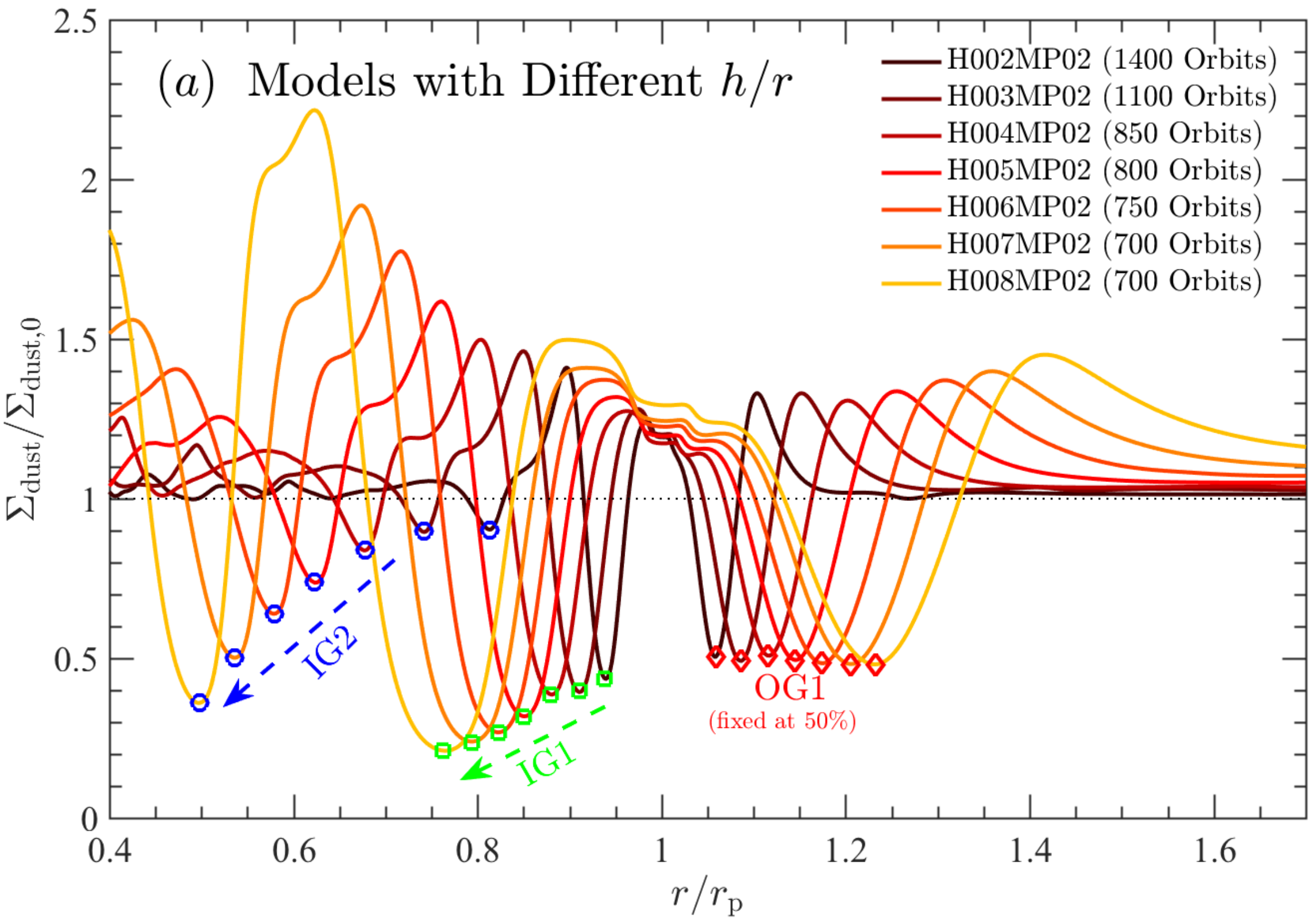}
\includegraphics[trim=0 0 0 0, clip,width=0.7\textwidth,angle=0]{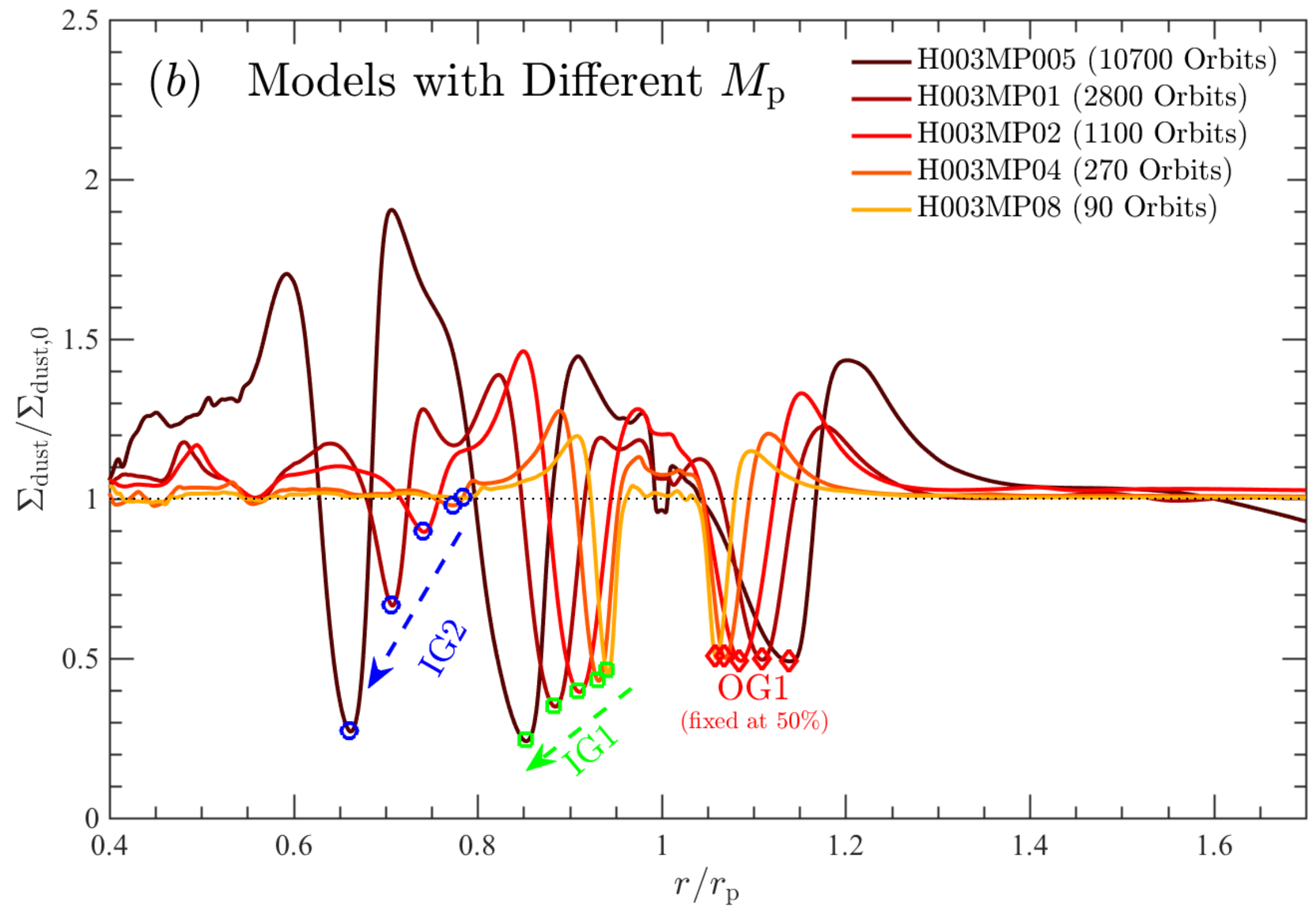}
\end{center}
\figcaption{{\it Top:} The radial profiles of $\sigmad/\Sigma_{\rm dust,0}$ in 7 models with $\mplanet=0.2\mth$ and successively larger $h/r$. {\it Bottom:} The radial profiles of $\sigmad/\Sigma_{\rm dust,0}$ in 5 models with $h/r=0.03$ and $\mplanet$ successively doubled. All models are shown at the time when OG1 reaches 50\% depletion. OG1, IG1, and IG2 are marked by red diamonds, green squares, and blue circles, respectively. As $h/r$ increases, or $\mplanet$ decreases, gaps move away from the planet and become more widely separated. In addition, gaps further from the planet become deeper relative to OG1 (dashed arrows). See \S\ref{sec:spacing} for details. 
\label{fig:rp}}
\end{figure*}

Figure~\ref{fig:rp} plots the radial profiles of $\sigmad/\Sigma_{\rm
  dust,0}$ in models with varying $h/r$ (top) and $\mplanet$
(bottom). The corresponding 2D dust maps can be found in the Appendix
Figure~\ref{fig:mphoverr}. All models are shown
when surface densities inside gaps
are $\sim$50\% of their original values.

As $h/r$ increases, or $\mplanet$ decreases, gaps move away from the planet and become more widely separated. Here we explain these trends in the framework of the weakly nonlinear density wave theory \citep{goodman01, rafikov02}, and provide empirical fitting functions for $r_{\rm OG1}$, $r_{\rm IG1}$, and $r_{\rm IG2}$.

Under the local approximation (constant $\sigmag$ and $h/r$ in the
planet's vicinity), \citet{goodman01} found that the
primary
density waves launched to either side of the planet's orbit
shock after traveling
a radial ``shocking length'' equal to
\begin{equation}
\lsh\approx0.8\left(\frac{\gamma+1}{12/5}\frac{\mplanet}{\mth}\right)^{-2/5}h,
\label{eq:lsh}
\end{equation}
where $\gamma$ is the adiabatic index ($\gamma=1$ for isothermal
gas). The shocks mark the onset of wave dissipation and gap opening in
gas. For planets with $\mplanet \sim 0.1$--$0.5\mth$, $\lsh\sim
1$--$2h$. We expect the shock locations $\rp\pm\lsh$ to mark the locations of dust
gaps IG1 and OG1, though the identification is not
perfect --- the shock
locations correspond to the edges of gas gaps, while dust gaps have finite
radial widths, and have different profiles from gas gaps. Another
issue is that Eqn.~(\ref{eq:lsh}) is only exact in the limit $\mplanet\ll\mth$. 

\begin{figure*}
\begin{center}
\includegraphics[trim=0 0 0 0, clip,width=0.49\textwidth,angle=0]{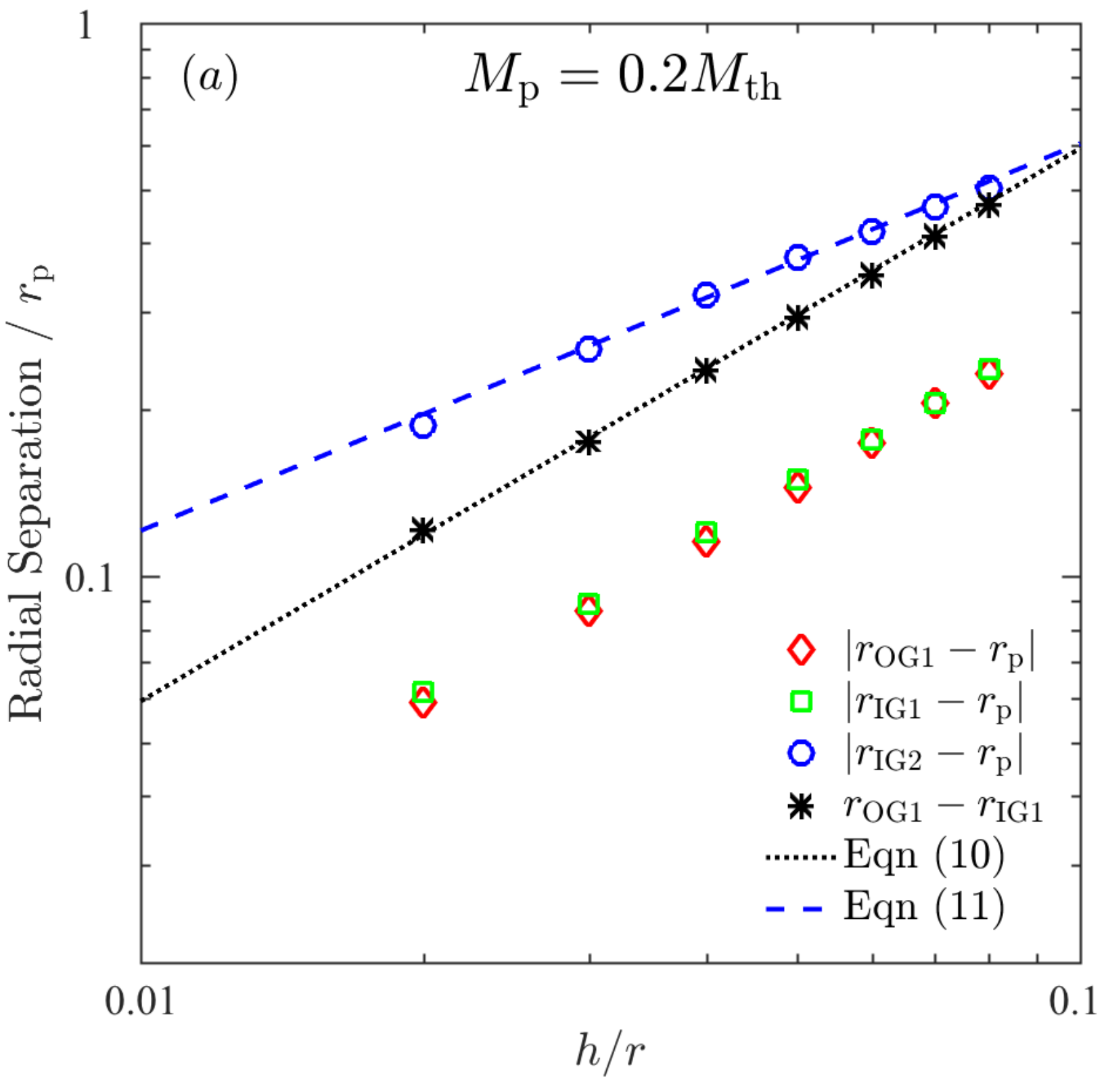}
\includegraphics[trim=0 0 0 0, clip,width=0.49\textwidth,angle=0]{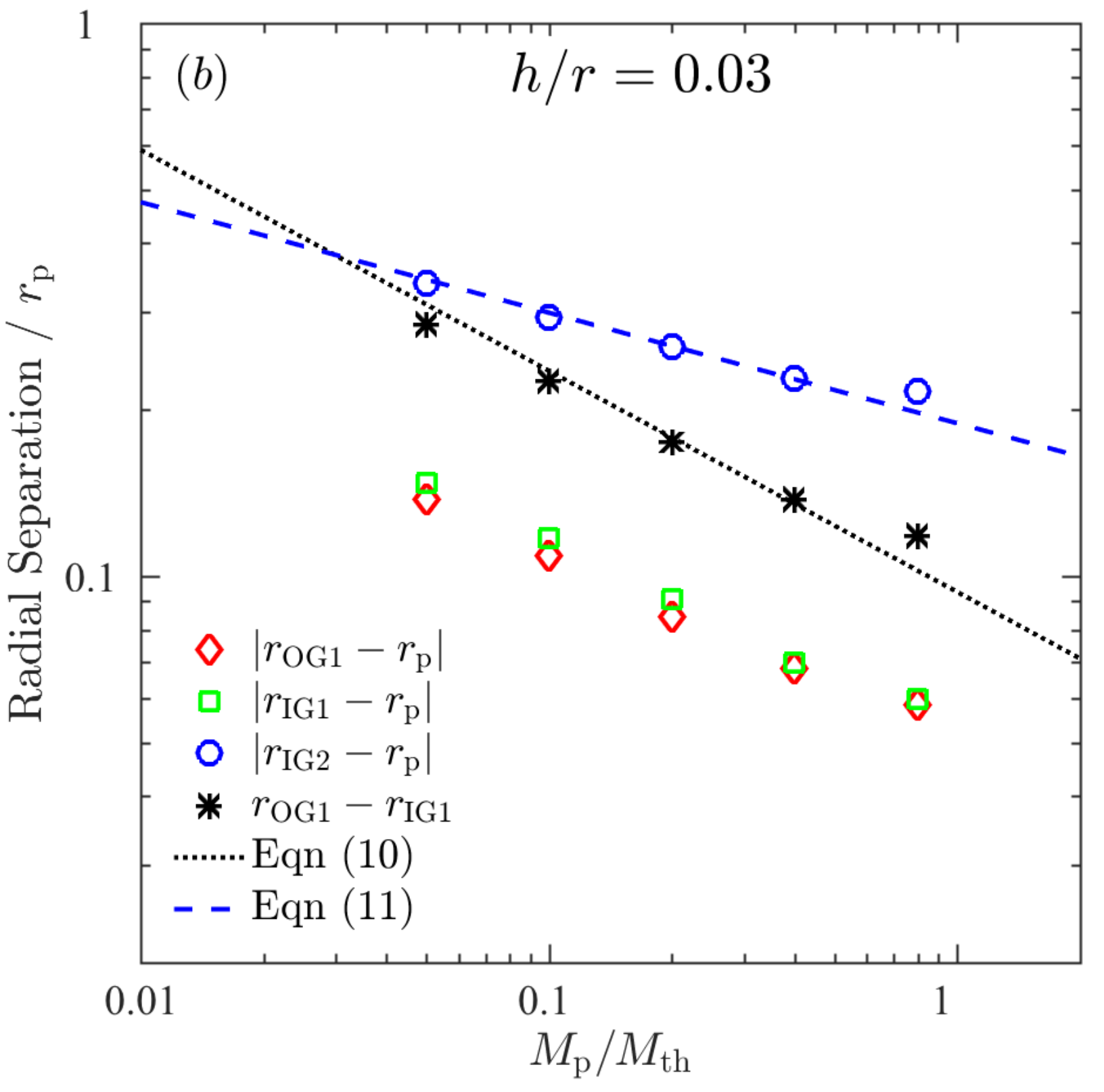}
\end{center}
\figcaption{{\it Left}: Gap spacings as a function of $h/r$ for the 7 models in Figure~\ref{fig:rp}($a$). The locations of OG1, IG1, and IG2 are plotted ($| r_{\rm gap}$$-$$\rp |$; symbols are the same as in Figure~\ref{fig:rp}), in addition to the distance between OG1 and IG1, $r_{\rm OG1}-r_{\rm IG1}$ (black stars). The dotted and dashed lines are the fitting functions~(\ref{eq:fit:og1ig1}) and (\ref{eq:fit:ig1ig2}), respectively. {\it Right}: Similar to ($a$), but for gap spacings as a function of $\mplanet$ for the 5 models in Figure~\ref{fig:rp}($b$). See \S\ref{sec:spacing} for details.
\label{fig:spacing}}
\end{figure*}

Nevertheless, we find the scaling relations implied by
Eqn.~(\ref{eq:lsh}) --- $\lsh\propto h$ and
$\lsh\propto\mplanet^{-2/5}$ --- fit the behaviors of $r_{\rm
  OG1}$$-$$\rp$ and $\rp$$-$$r_{\rm IG1}$ well, as shown in
Figure~\ref{fig:spacing}. Consequently, $r_{\rm OG1}$$-$$r_{\rm IG1}$,
a.k.a.~the ``double gap'' separation, obeys the same
scalings. Empirically we find that
\begin{equation}
\frac{r_{\rm OG1}-r_{\rm IG1}}{\rp} \approx2.9\left(\frac{\gamma+1}{12/5}\frac{\mplanet}{\mth}\right)^{-2/5} \left(\frac{h}{r}\right).
\label{eq:fit:og1ig1}
\end{equation}
We expect the agreement to deteriorate as $\mplanet$ approaches $\mth$.

The gaps inward of IG1 appear opened by the dissipation of additional
(secondary, tertiary, etc.) density waves excited by the planet
\citep{bae17}. No theoretical calculations quantifying how these waves
dissipate have been carried out. We find that $\rp-r_{\rm IG2}$ does
not scale with $\mplanet$ and $h/r$ as in
Eqns.~(\ref{eq:lsh})--(\ref{eq:fit:og1ig1}); instead, the
location of IG2 can be fitted by:
\begin{equation}
1-\frac{r_{\rm IG2}}{\rp} \approx 2.2 \left(\frac{\mplanet}{\mth}\right)^{-0.2} \left(\frac{h}{r}\right)^{0.7}.
\label{eq:fit:ig1ig2}
\end{equation}
Compared with those of the double gap, the dependences of
$\rp$$-$$r_{\rm IG2}$ on both $h/r$ and $\mplanet$ are weaker. 

Empirically fitting the locations of gaps further inward of IG2 is
possible, but only a small number of models robustly reveal such gaps
and so we have not performed this exercise. 
 
\subsection{The Number and Depths of Gaps}\label{sec:depth}

\begin{figure*}
\begin{center}
\includegraphics[trim=0 0 0 0, clip,width=0.49\textwidth,angle=0]{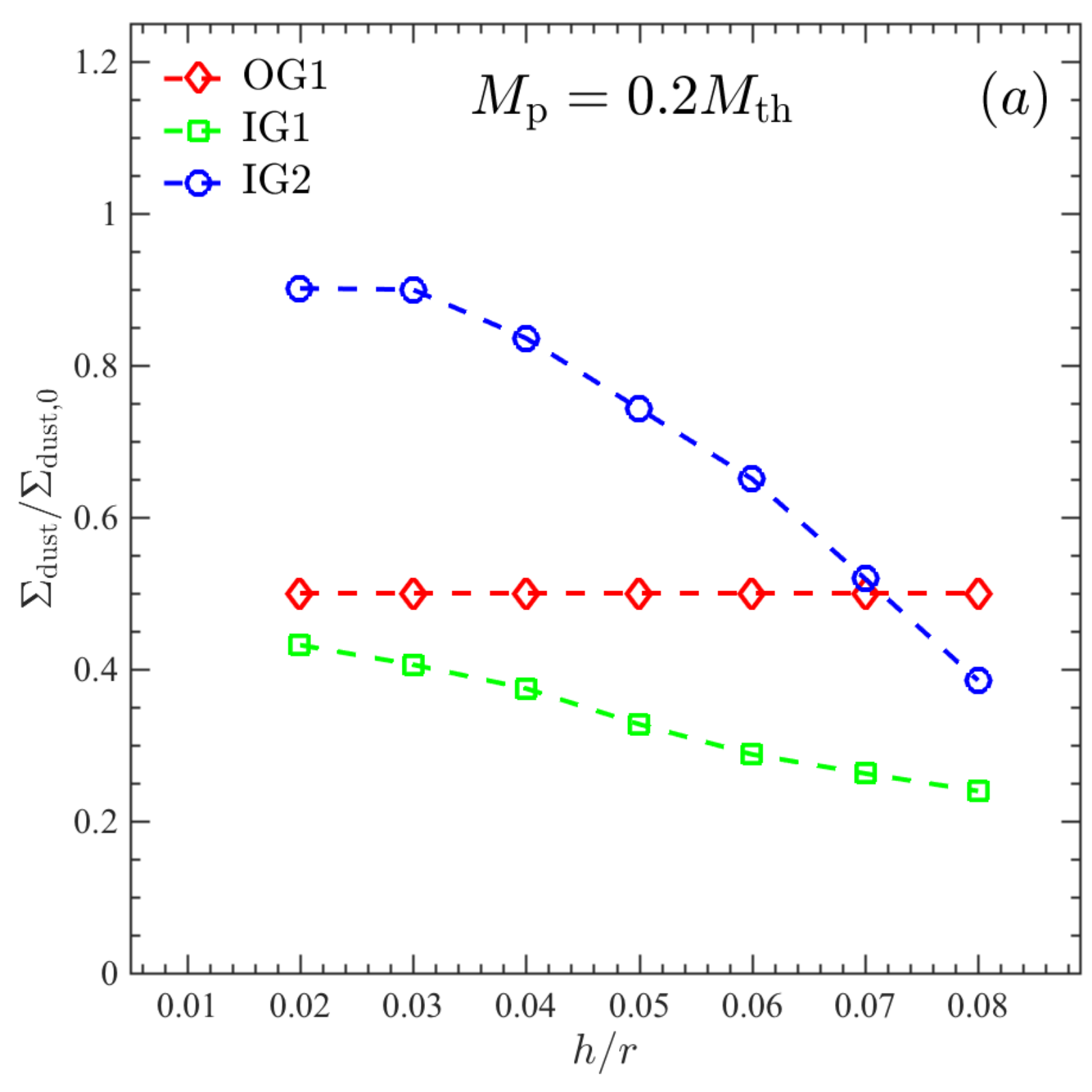}
\includegraphics[trim=0 0 0 0, clip,width=0.49\textwidth,angle=0]{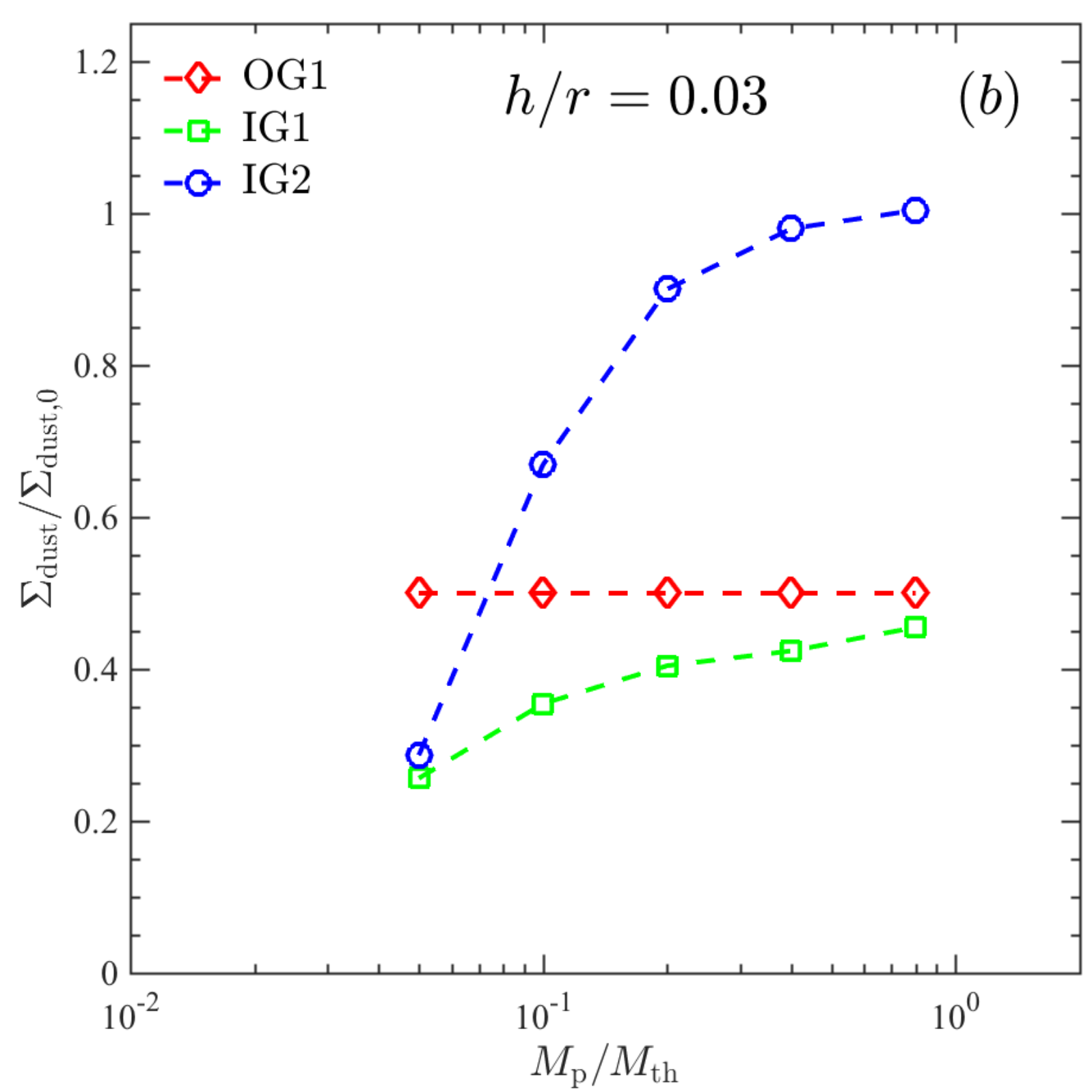}
\end{center}
\figcaption{{\it Left}: The dependence of gap depth on $h/r$, showing $\sigmad/\Sigma_{\rm dust,0}$ at OG1/IG1/IG2 for the 7 models in Figures~\ref{fig:rp}($a$) and \ref{fig:spacing}($a$) (normalized relative to OG1 with $\sigmad/\Sigma_{\rm dust,0}=0.5$). {\it Right}: The dependence of gap depth on $\mplanet$, showing the 5 models in Figures~\ref{fig:rp}($b$) and \ref{fig:spacing}($b$). As $h/r$ increases, or $\mplanet$ decreases, the gaps inside $\rp$ deepen relative to OG1 (more so for the inner ones). See \S\ref{sec:depth} for details.
\label{fig:depth}}
\end{figure*}

In contrast to gap spacings, the total number of gaps
opened by a planet evolves more strongly with time. As they deepen,
more gaps become visible (e.g., Figure~\ref{fig:time}).
How fast a gap deepens ($d\sigmad/dt$) depends on a variety
of factors. A few general observations:
\begin{enumerate}
\item Different gaps deplete at different rates.
For example, while in Model H003MP02 OG1 and IG1 deepen at roughly the
same rates, gaps farther from the planet's orbit deplete more
slowly (Figure~\ref{fig:time}). 
\item At fixed $\mplanet/\mth$, 
gaps open faster in disks with higher $h/r$. 
Figure~\ref{fig:rp}($a$)
shows that in such disks it takes less time to deplete OG1,
and IG1 and IG2 deepen even faster. 
Figure~\ref{fig:depth}($a$) shows
$\sigmad/\Sigma_{\rm dust,0}$ at OG1, IG1, and IG2 for these models.
\item More massive planets open gaps faster. Figure~\ref{fig:rp}($b$)
  shows that a planet 10$\times$ more massive depletes OG1
  $\sim$100$\times$ faster. This effect is weaker for gaps 
farther from the planet. Figure~\ref{fig:depth}($b$) shows
$\sigmad/\Sigma_{\rm dust,0}$ at OG1, IG1, and IG2 for the same models in Figure~\ref{fig:rp}($b$).
\end{enumerate}
Gaps at small radii are difficult to count in both models and
observations
because of limited resolution. 
We list the number of dust gaps
between 0.4$\rp$ and OG1 (inclusive; OG2 is never prominent)
in Table~\ref{tab:models} (last column).
A gap is counted as long as its $d\sigmad/dt$ is consistently negative
(i.e., no minimum threshold in $\sigmad/\Sigma_{\rm dust,0}$ is
imposed).
Not all gaps are deep enough to be detectable in realistic observations. 

In general, a planet opens more gaps in disks with lower $h/r$ ---
gaps in such a disk are 
narrower, 
allowing more gaps to 
fit within a given radius range (e.g., $0.4\rp\leq r\leq r_{\rm  OG1}$).
The trend is less clear with 
planet mass. There are two competing factors. \citet{bae18simulation} 
discerned more density waves with decreasing
$\mplanet$, potentially resulting in more gaps. On the other hand,
gaps opened by a less massive planet are more widely separated,
decreasing the maximum number of gaps within a given radius
interval. It is unclear which effect dominates.
Tentatively, models
H003MP01--H003MP08 each have five gaps at $r\ge0.4\rp$, suggesting
that the number of gaps may be insensitive to $\mplanet$ (note that Model H003MP005 has only three gaps).


\section{Comparison to Real Disks}\label{sec:applications}

In this section, we make a quick comparison between our
models and observed disks,
focussing on those having three
or more narrow gaps in mm continuum
emission (HL Tau, TW Hya, and HD 163296). 
These comparisons do not involve detailed fits. Rather, the 
models are taken directly from our parameter survey grid, and 
are not specifically tailored to the observations. 
We focus on gap locations and not on gap depths for
reasons discussed in \S\ref{sec:comments}.
We begin by picking $h/r$ consistent with disk 
midplane temperatures, $\tmidplane$, as derived from
the literature. We then identify a pair of gaps as
the ``double gap''. For HL Tau and TW Hya,
this is the most compact pair, and for HD 163296,
it is the pair farthest from the star.
The midline of the double gap locates the planet position
$\rp$. We then estimate $\mplanet$ using 
Eqn.~(\ref{eq:fit:og1ig1}), selecting the one model
from our grid that appears to best reproduce the double
gap locations. A non-trivial test of the model is to see
whether it can reproduce the locations of observed
gaps interior to the double gap.

\begin{figure*}
\begin{center}
\includegraphics[trim=0 0 0 0, clip,width=\textwidth,angle=0]{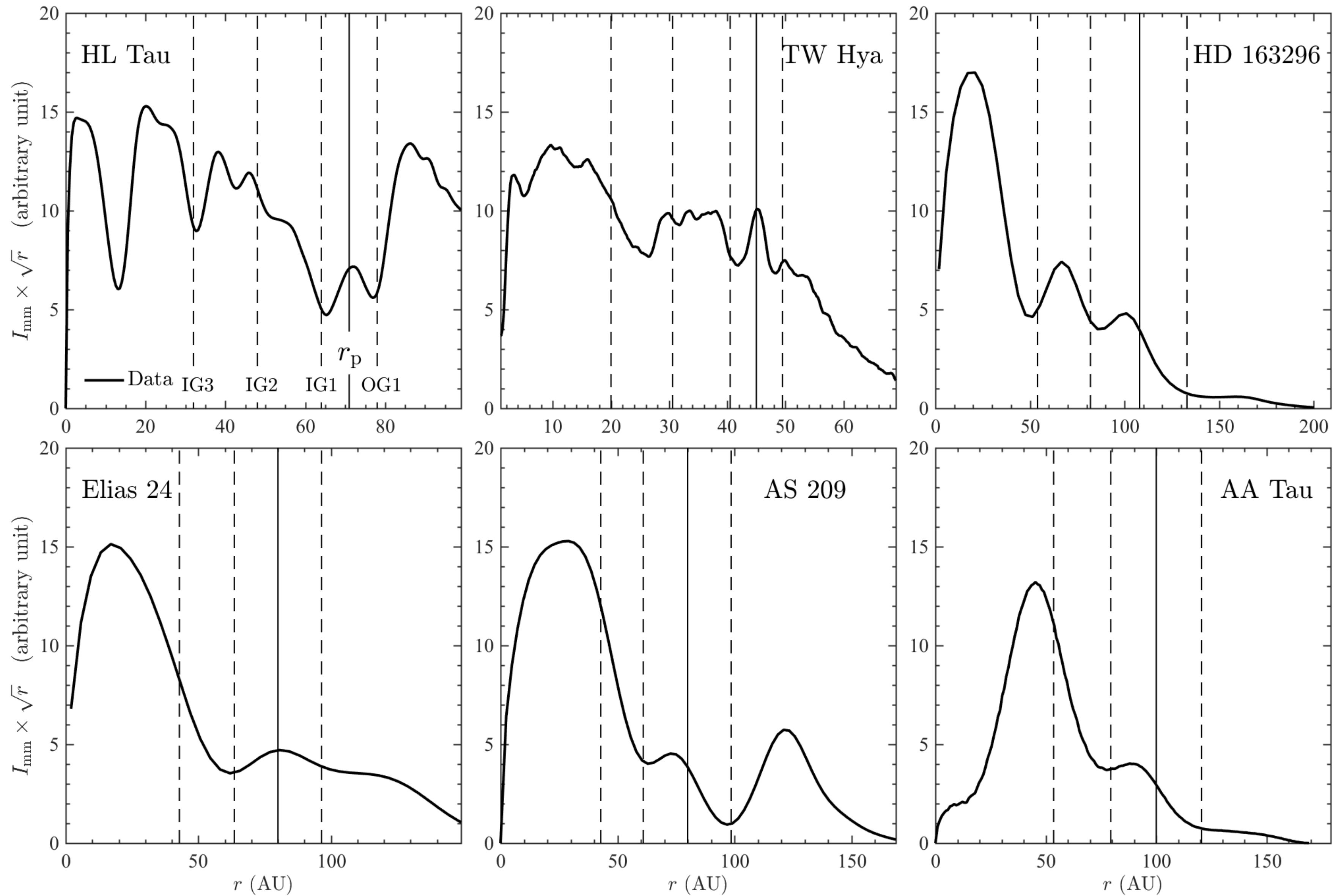}
\end{center}
\figcaption{Radial profiles of surface brightness in mm continuum emission ($I_{\rm mm}$; solid curve) in the HL Tau \citep{brogan15}, TW Hya \citep{andrews16}, HD 163296 \citep{isella16hd163296}, Elias 24 \citep{dipierro18}, AS 209 \citep{fedele18}, and AA Tau disks \citep{loomis17}. These profiles are azimuthally averaged (except in HL Tau, where the profile is measured along the disk major axis), and scaled by $\sqrt{r}$ to suppress (partially) the dependence of $\imm$ on dust temperature.
The vertical solid line in each panel marks the location of the planet $\rp$ in our models, and the vertical dashed lines mark the locations of all modelled gaps at $r\ge0.4\rp$ (regardless of depth). Models are selected from our parameter 
grid such
that the locations of gaps
IG1 and OG1 (comprising the ``double gap'')
are approximately reproduced in each case.
In HD 163296, we are able to match a third gap IG2.
In TW Hya, none
of the predicted gaps
(apart from IG1 and OG1, which fit by construction)
appear to match.
In HL Tau, IG3 and IG2 may correspond to observed features, while
the observed gaps at 14 and 42 AU (D1 and D3)
are not reproduced (D1 lies too close to the inner boundary
of our simulation to be accurately modeled; see
Figure \ref{fig:resolutionrp}).
In Elias 24, AS 209, and AA Tau, we introduce simple models to explain the two gaps revealed by existing observations, assuming they compose the double gap and $\mplanet=0.2\mth$. The location of a third gap IG2 in the models (the innermost vertical dashed line) is labeled too. See \S\ref{sec:applications} for details.
\label{fig:realdisks}}
\end{figure*}

The observed surface brightness radial profiles $\imm(r)$ are 
shown in Figure~\ref{fig:realdisks}, with modelled planet and 
gap locations indicated. We mark all modelled gaps at 
$r\ge0.4\rp$ regardless of their depths, but do not
show the model surface brightness radial profile, again
because our intent is not to fit gap depths.
We attempt to remove 
the dependence of $\imm$ on dust temperature by scaling 
$\imm$ by $\sqrt{r}$.

\begin{figure*}
\begin{center}
\includegraphics[trim=0 0 0 0, clip,width=0.75\textwidth,angle=0]{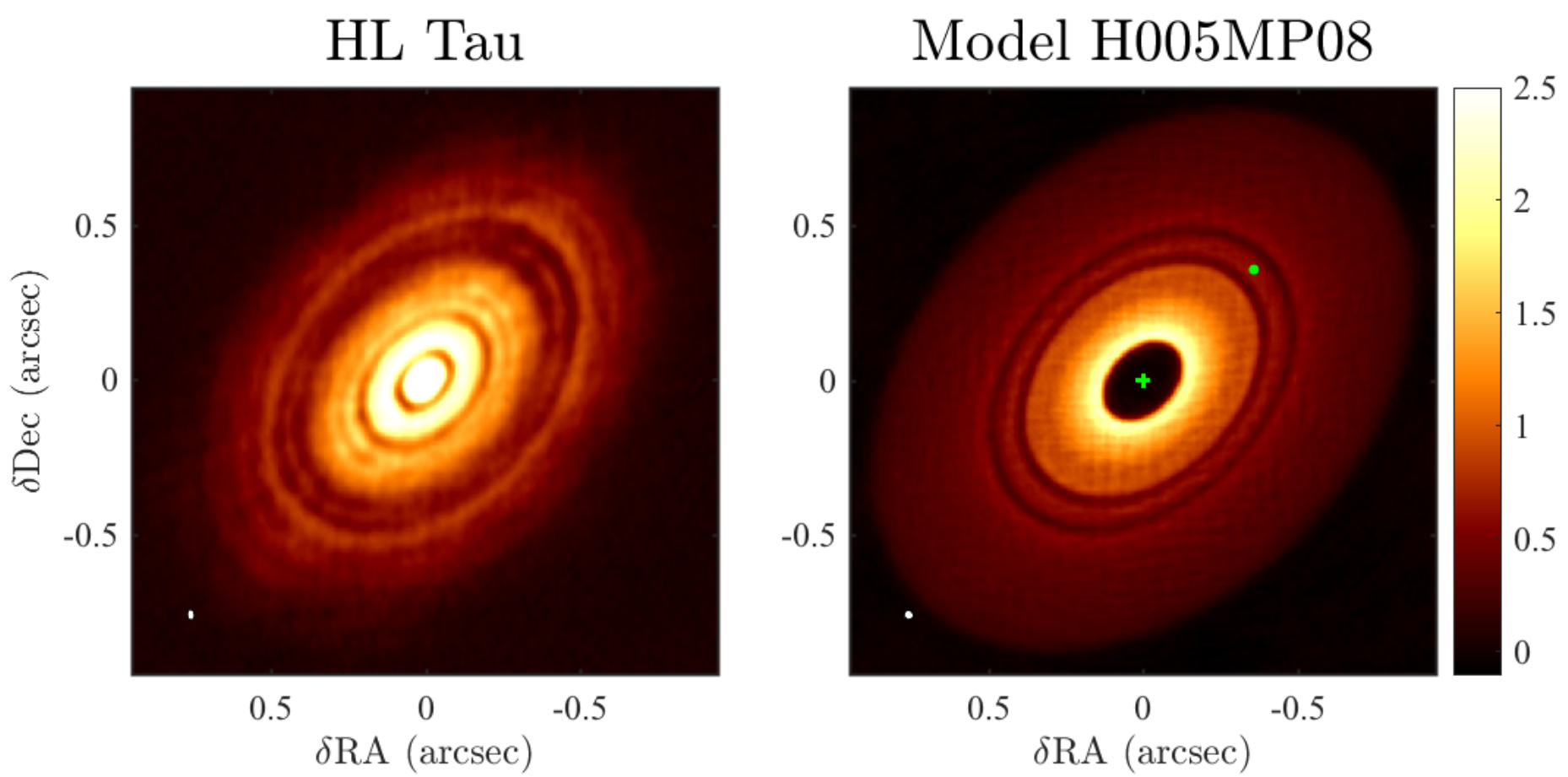}
\includegraphics[trim=0 0 0 0, clip,width=0.75\textwidth,angle=0]{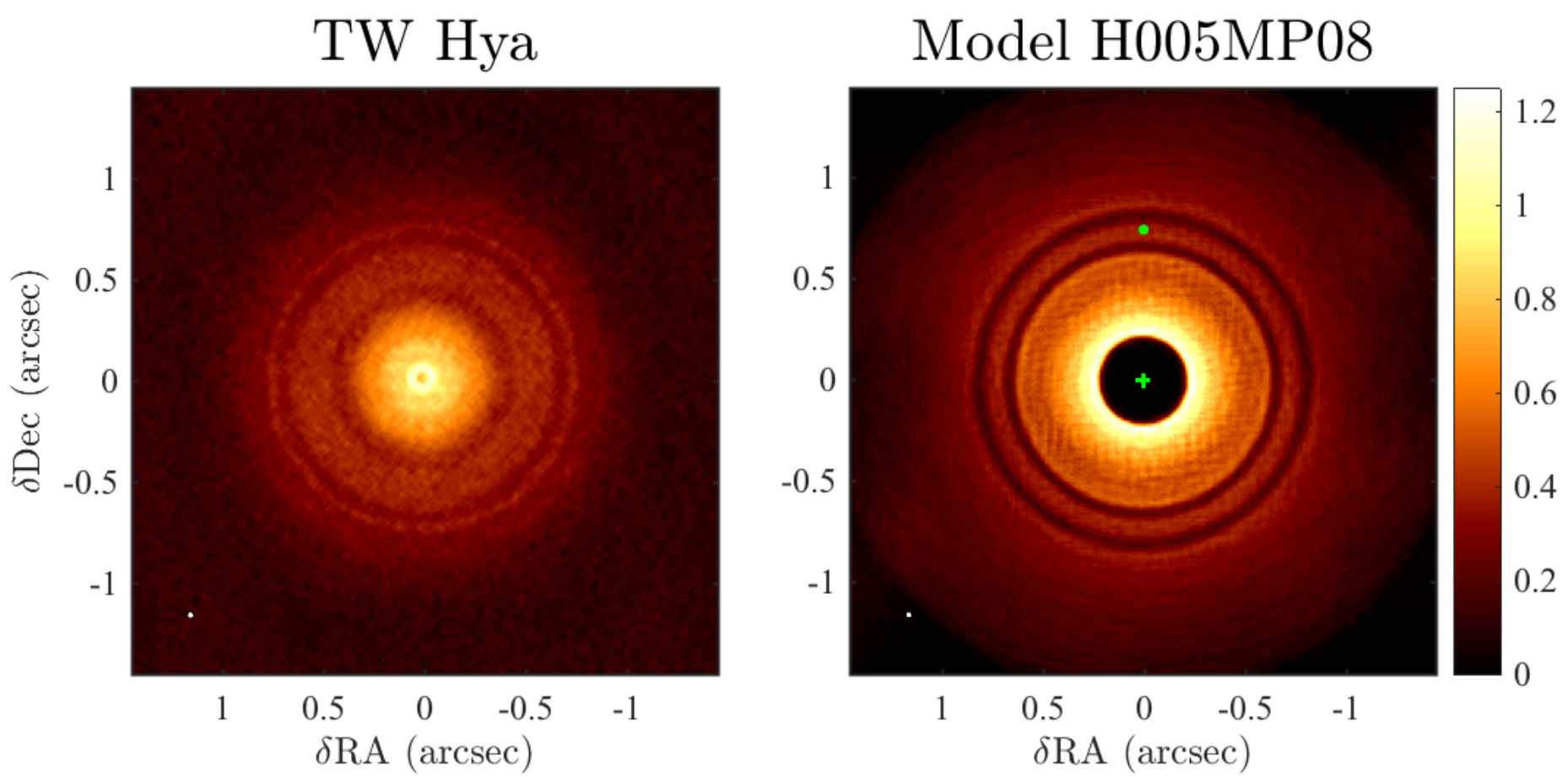}
\includegraphics[trim=0 0 0 0, clip,width=0.75\textwidth,angle=0]{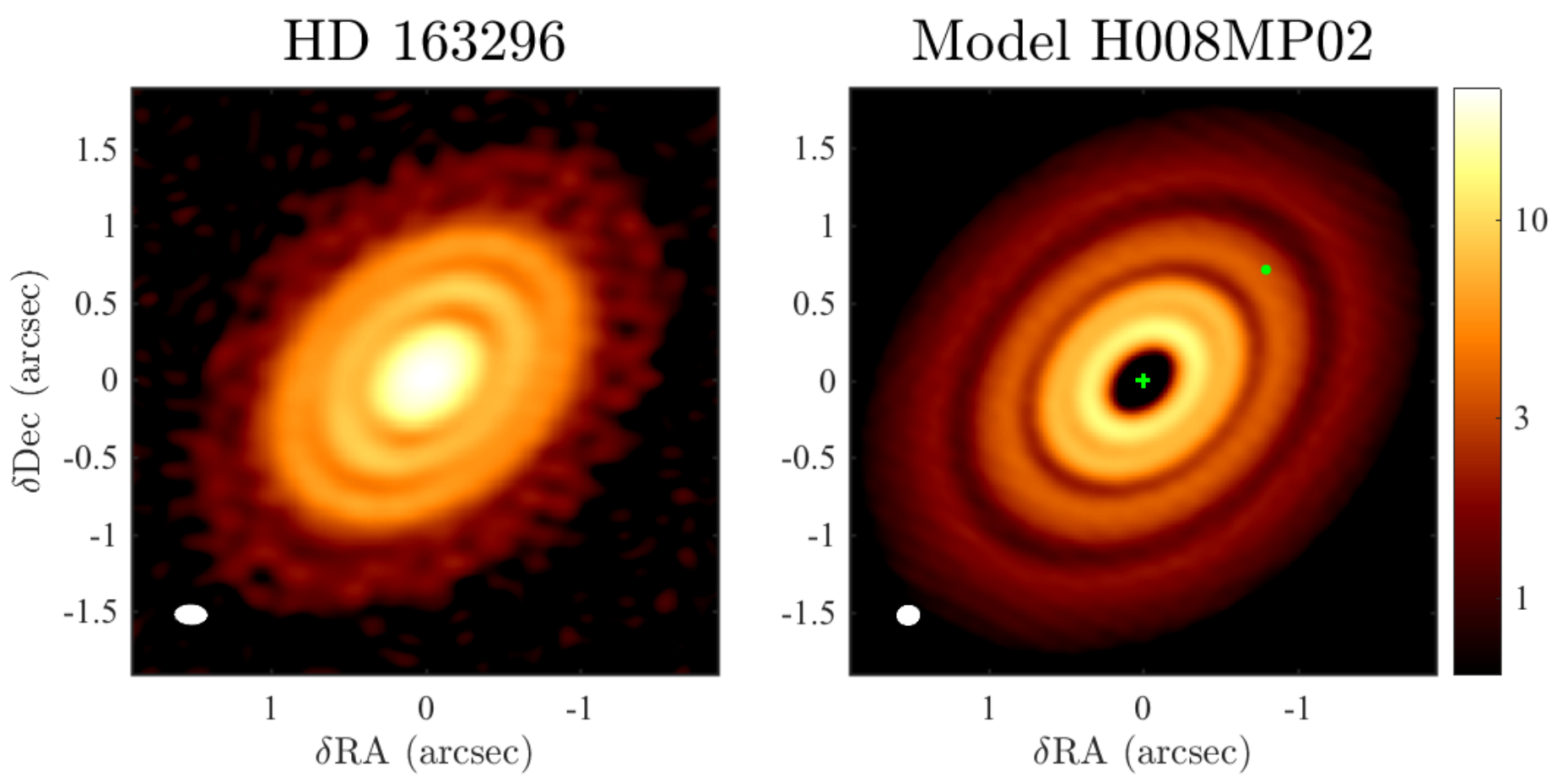}
\end{center}
\figcaption{\small {\it Top row:} Comparison between ALMA observations of HL Tau \citep{brogan15} and Model H005MP08. The model has been scaled such that $\rp=0\farcs51$ (71 AU);  $\mplanet=57\me$; the elapsed time equals 110 orbits (0.05 Myr); its total flux density is 1 Jy; and the viewing geometry matches the HL Tau disk. The green plus and dot symbol mark the star and the planet, respectively. The beam is indicated in the bottom left corner. The innermost model disk regions are manually masked out, and the color stretch is linear. {\it Middle row:} Similar to the top row, but for TW Hya \citep{andrews16}. The model (also scaled from H005MP08) has a 29$\me$ planet at 0\farcs75 (45 AU). {\it Bottom row:} Similar to the top row, but for HD 163296 \citep{isella16hd163296}. The underlying
model, H008MP02, has a 65$\me$ planet at 1\farcs07 (108 AU), and has been run for 700 orbits (0.6 Myr). The color stretch for this row only is logarithmic. The locations of some but not all modelled gaps coincide with those of observed gaps. See \S\ref{sec:applications} for details.
\label{fig:sidebyside}}
\end{figure*}

Figure~\ref{fig:sidebyside} shows side-by-side comparisons of observations with synthetic model ALMA continuum images, produced using radiative transfer simulations using the \citet{whitney13} code. The procedures are described in \citet[see also \citealt{dong17doublegap}]{dong15gap}. Briefly, 2D hydro disk structures are inflated into 3D, assuming hydrostatic equilibrium with each model's $h/r$. Starlight is reprocessed to longer wavelengths at the disk surface by a population of ``small'' (sub-$\mu$m-sized) interstellar medium dust (\citealt{kim94}). These small grains are assumed to be well-coupled to gas with a volumetric mass density equal to 10$^{-2}$ that of the gas. A population of ``big'' dust
particles (representing the second fluid in hydro simulations)
is assumed to have settled to the midplane with a Gaussian vertical density profile and a scale height equal to $h_{\rm dust}=0.1h$, in line with $h_{\rm dust}$ estimated in real disks \citep[e.g., HL Tau, $\sim$1\% at 100 AU;][]{pinte16}. The optical properties of the small dust can be found in \citet[Fig.~2]{dong12cavity}. The big dust is assumed to have an opacity $\kappa=3.5$~cm$^2$~g$^{-1}$ at 870 $\micron$. For the central source we assume a 1$\msun$ pre-main-sequence star with $R_\star=2.4R_\odot$ and $T_\star=4400$K. Radiative transfer simulations produce synthetic images of continuum emission at 870 $\micron$, which are processed using the \texttt{simobserve} and \texttt{simanalyze} tools in Common Astronomy Software Applications (CASA) to simulate ALMA images with 4-hour integration times and default settings.

\paragraph{HL Tau}
Millimeter continuum observations reveal four major gaps at 14, 32, 65, and 77 AU (labelled D1, D2, D5, and D6 in \citealt{brogan15}; locations are adopted from the \citealt{akiyama16hltau} re-analysis), and shallower gaps D3, D4, and D7 at 42, 50, and 91 AU. We interpret the small fractional separation of D5 and D6 ($\Delta_{\rm D5, D6}=0.17$) and their similar depths as  signatures of a double gap (IG1 and OG1). At 70 AU, \citet{kwon11} and \citet{jin16} found  $\tmidplane\sim13$ K and 25 K, respectively, by performing radiative transfer modeling of continuum emission.
We adopt $h/r=0.05$, corresponding to $\tmidplane=17$ K ($M_{\rm HL\ Tau}=1.7\msun$, \citealt{pinte16}). Eqn.~(\ref{eq:fit:og1ig1}) then yields $\mplanet \simeq 0.8\mth$ (57$\me$). The corresponding Model H005MP08 has four gaps at $r\ge0.4\rp$: $r_{\rm IG3}$=32 AU, $r_{\rm IG2}$=48 AU, $r_{\rm IG1}$=64 AU, and $r_{\rm OG1}$=78 AU for $\rp=71$ AU. 
Figure~\ref{fig:sidebyside} (top) shows an image comparison. 
These four gaps roughly coincide with D2, D4, D5, and D6, respectively. D1, D3, and D7 in HL Tau find no ready correspondence with our model. \citet{bae17} fit D1, D2, D5, and D6 using a $30\me$ planet at 69 AU with $h/r=0.074$. Our model better reproduces the observed double gap ($\Delta_{\rm D5, D6} \simeq 0.28$ according to \citealt{bae17}),
mainly due to our lower $h/r$.

\paragraph{TW Hya}
Continuum observations at 0.87 mm reveal three gaps at 25, 41,
and 49 AU (G1, G2, and G3, \citealt{andrews16}; these distances
have been rescaled using the \citealt{gaia18} distance to
TW Hya of 60 pc). Gap G3
is better revealed in the spectral index map than in the emission map (\citealt{huang18}). We identify G2 and G3 as a double
gap, with $\Delta_{\rm G2,G3}=0.17$.
\citet{andrews16} found $\tmidplane\sim15$ K at 45 AU
using radiative transfer modeling, corresponding to $h/r\simeq 0.05$ 
($M_{\rm TW\ Hya}=0.88\msun$, \citealt{huang18}). 
These parameters are essentially identical to those in HL Tau; 
accordingly, Model H005MP08 is appropriate for TW Hya as well.
A planet with $\mplanet= 0.8 \mth \, (29\me)$, located at $\rp = 45$ AU, opens four gaps at $r\ge0.4\rp$: $r_{\rm IG3}=20$ AU, 
$r_{\rm IG2}=30$ AU, $r_{\rm IG1}=41$ AU, and $r_{\rm OG1}=50$ AU. 
Figure~\ref{fig:sidebyside} (middle) shows an image comparison. The model roughly reproduces the spacing of the observed double gap, while G1 finds no ready correspondence with our model.

\paragraph{HD 163296}
Continuum observations at 1.3 mm reveal three gaps at
50, 83, and 133 AU (G1, G2, and G3, \citealt{isella16hd163296};
these distances have been rescaled using the \citealt{gaia18} 
distance to the star of 101 pc). 
Their locations
roughly match the three gaps IG2, IG1, and OG1 at $r\ge0.4\rp$ opened by a planet with $\mplanet = 0.2 \mth = 65\me$ ($M_{\rm HD\ 103296}=1.9\msun$; \citealt{fairlamb15}) in Model H008MP02: $r_{\rm IG2}$=54 AU, $r_{\rm IG1}$=82 AU, and
$r_{\rm OG1}$=133 AU for $\rp=108$ AU.
Figure~\ref{fig:sidebyside} (bottom) shows an image comparison.
The adopted $h/r = 0.08$ is roughly
consistent with $\tmidplane\sim20$~K at $\rp$
as inferred by \citet{rosenfeld13hd163296} and \citet{liu18}. Note that our identification of the double gap, G2+G3, is ambiguous --- it is also possible to identify G1+G2 as the double gap and find a corresponding model, though in that case G3 cannot be accounted for.
\\

A few systems have been found to host two dust gaps in existing observations (e.g., AS 209, \citealt{fedele18}; AA Tau, \citealt{loomis17}; V1094 Sco, \citealt{vanterwisga18}; and Elias 24, \citealt{dipierro18}). Interpreting each pair as a double gap, we can infer the locations of possible additional, as yet unseen inner gaps. Taking $\mplanet=0.2\mth$, the two gaps in the Elias 24, AS 209, and AA Tau disks coincide with the double gap opened by a 23$\me$ planet at 80 AU with $h/r=0.07$, a 31$\me$ planet at 80 AU with $h/r=0.08$, and a 19$\me$ planet at 100 AU with $h/r=0.07$, respectively (all distances rescaled using data from the \citealt{gaia18}; $M_{\rm Elias\ 24}=1\msun$, \citealt{wilking05}; $M_{\rm AS\ 209}=0.9\msun$, \citealt{tazzari16}; $M_{\rm AA\ Tau}=0.85\msun$, \citealt{loomis17}). The location of IG2 in each model is marked in Figure~\ref{fig:realdisks}. In every case, IG2 falls where $\imm$ has a steep radial gradient, hindering the detection of possible gaps to date. 
We emphasize that these models are not unique, and they may need to be updated if future observations reveal additional sub-structures.
The two gaps in the V1094 Sco disk (not shown) are too far apart to be explained by models with $h/r\leq0.08$.

In closing this section, we note that gas gaps in our models are shallow ($\delta\sigmag/\Sigma_{\rm gas, 0}\sim1-10\%$),
and so are not expected to be easily detectable in near-infrared (NIR) imaging, which traces (sub-)$\micron$-sized dust typically well-coupled to gas \citep{dong17gap}. This expectation generally accords with observations. In HD 163296, a broad gap is present at $\sim$0$\farcs2-0\farcs5$ in scattered light \citep{garufi14, monnier17}, but no counterparts to the three narrow ALMA dust gaps at $r \gtrsim 0\farcs5$ are evident
in the NIR (however note the detection of line intensity depressions in CO isotopologues at locations similar to the dust gaps; \citealt{isella16hd163296}). In AA Tau, no scattered light gaps are seen at the locations of mm-dust gaps \citep{cox13}. In TW Hya, G2 and G3 are not detected in scattered light; however, G1 is revealed at a slightly smaller radius \citep{akiyama15, rapson15twhya, vanboekel17, debes17}, indicating a large depletion in gas. High fidelity NIR disk imaging is not available for HL Tau (but note the tentative detection of gas gaps in HCO$^+$, \citealt{yen16}), AS 209, and Elias 24. 

\section{Discussion}\label{sec:discussions}

\subsection{The Observed Double Gap}\label{sec:doublegap}

To explain the double gap in the HL Tau and TW Hya disks, two mechanisms have been proposed: dust sintering and evolution at snowlines \citep{zhang15, okuzumi16}, and planet-disk interactions along the lines of our models. The former needs the snowlines of two relevant volatiles to be rather close (separation/radius ratio $\sim17\%$), while the latter posits a sub-Saturn planet at dozens of AU. While accurate assessments of snowline locations are needed to test the former hypothesis, the slow variation of disk temperature with radius may be difficult to make work.

Recently, an eccentric double gap with $e\sim0.1$ was tentatively discovered in the MWC 758 disk \citep{dong18mwc758}. A non-zero eccentricity may be compatible with a planet on an eccentric orbit, but is not natural to the snowline mechanism, which has no explicit azimuthal dependence. 

\subsection{Low Viscosity and the Rossby Wave Instability}\label{sec:lowalpha}

\begin{figure*}
\begin{center}
\includegraphics[trim=0 0 0 0, clip,width=0.9\textwidth,angle=0]{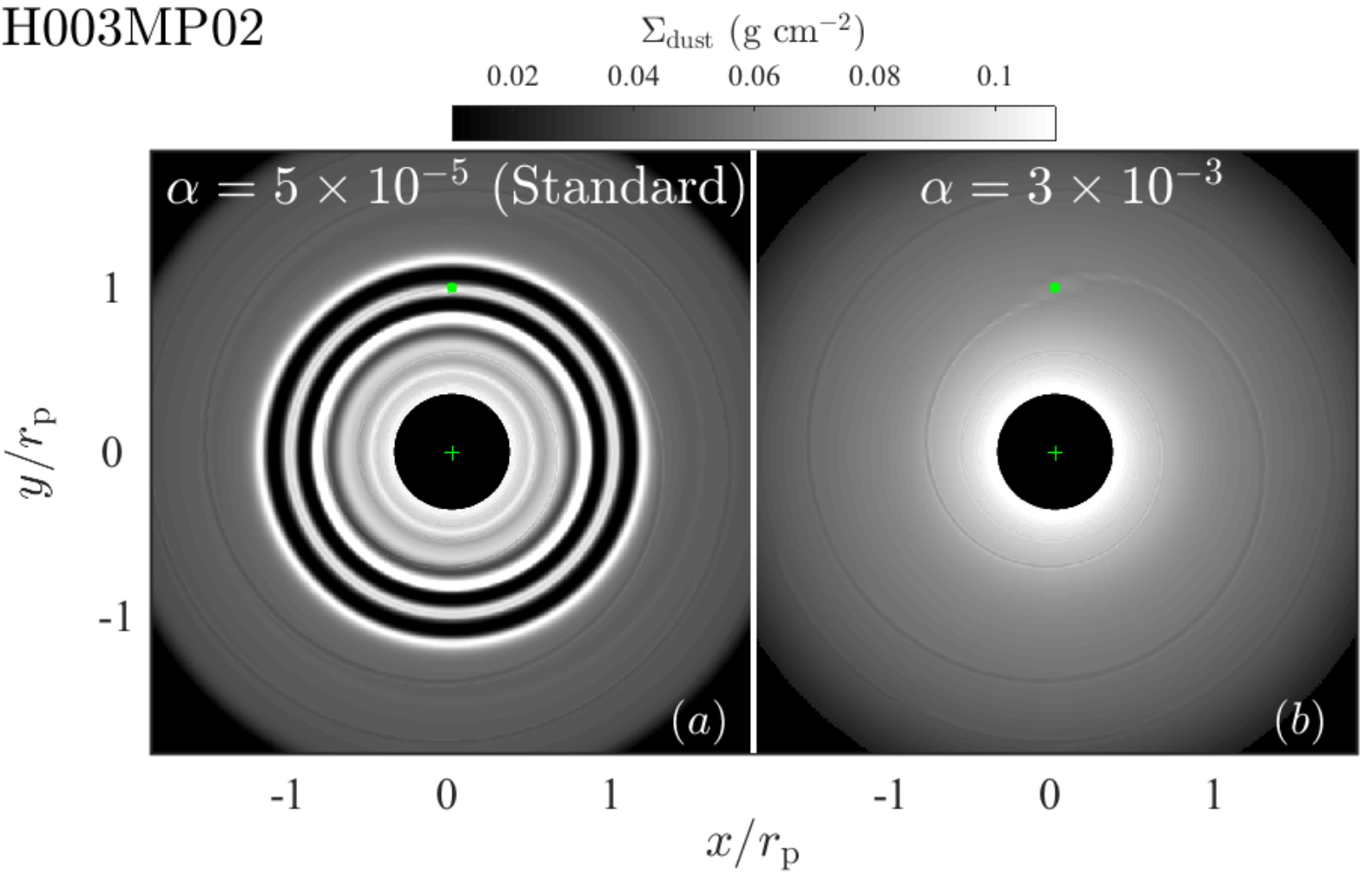}
\end{center}
\figcaption{Two runs of Model H004MP02 at 1600 orbits with $\alpha=5\times10^{-5}$ (left) and $\alpha=3\times10^{-3}$ (right). A low viscosity is required for a sub-thermal mass planet to open multiple dust gaps. See \S\ref{sec:lowalpha} for details.
\label{fig:alpha}}
\end{figure*}

A low viscosity ($\alpha\lesssim10^{-4}$) at the disk midplane 
(where $\sim$mm-sized dust particles are expected to settle) 
is required for a planet with $\mplanet\lesssim\mth$ to open multiple gaps. Figure~\ref{fig:alpha} compares two runs of Model H003MP02 with $\alpha=5\times10^{-5}$ (left; standard $\alpha$) and $3\times10^{-3}$ (right). No dust gaps are opened in the latter, while five gaps are opened in the former. 

Evidence for little-to-no turbulence at the disk midplane at tens of AUs is accumulating both theoretically  \citep[e.g.,][]{perezbecker11td, bai13ad, bai15, schlaufman18} and observationally \citep[e.g.,][see also the discussion in \citealt{fung17}, final section]{pinte16, flaherty17}. If $\alpha\lesssim10^{-4}$ is common, narrow dust gaps should mostly appear in pairs or multiples.

\begin{figure}
\begin{center}
\includegraphics[trim=0 0 0 0, clip,width=0.45\textwidth,angle=0]{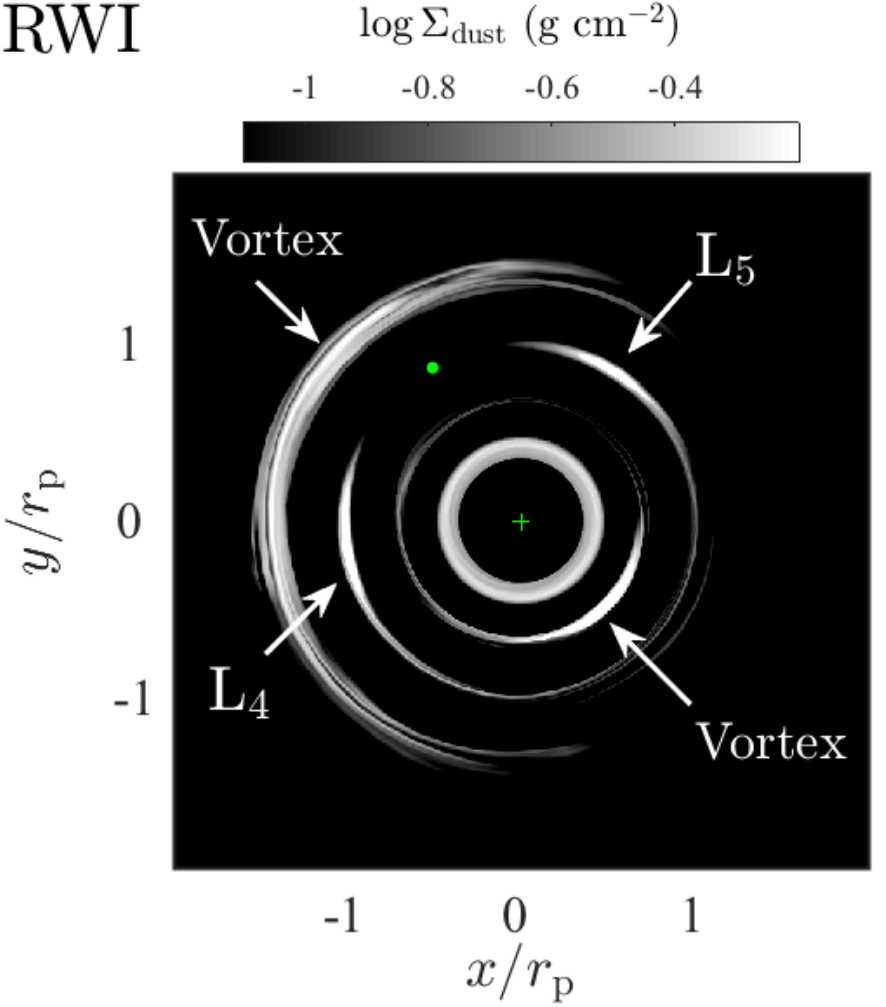}
\end{center}
\figcaption{The dust surface density map for Model RWI at 1200 orbits in log scale with an aggressive color stretch. The green dot and the green plus symbol mark the locations of the planet and the star, respectively. Two vortices triggered by the RWI form at the outer edge of OG1 and the inner edge of IG1. Two dust clumps at the triangular Lagrange points L$_4$ and L$_5$ are also present. In total there are 4 dust clumps in this disk with one planet. See \S\ref{sec:lowalpha} for details.
\label{fig:rwi}}
\end{figure}

In low viscosity environments, gas gaps sufficiently deep are prone to the Rossby wave instability \citep[RWI; e.g.,][]{li01, li05} at their edges. Gaps in this paper are generally too shallow to develop the RWI. Experiments (not shown) suggest that $\sigmag$ needs to be depleted by
$\gtrsim 50$\% to trigger the RWI when  $\alpha=5\times10^{-5}$. The RWI may develop at the edges of more than one gap, forming multiple dust-trapping vortices. Figure~\ref{fig:rwi} shows an example Model RWI, in which the RWI is triggered at both the inner edge of IG1 and the outer edge of OG1, forming two vortices. Dust are also collected at the triangular Lagrange points L$_4$ and L$_5$. In total, four dust clumps are produced.

\subsection{Implications for Planet Formation}\label{sec:implication}

\begin{figure}
\begin{center}
\includegraphics[trim=0 0 0 0, clip,width=0.45\textwidth,angle=0]{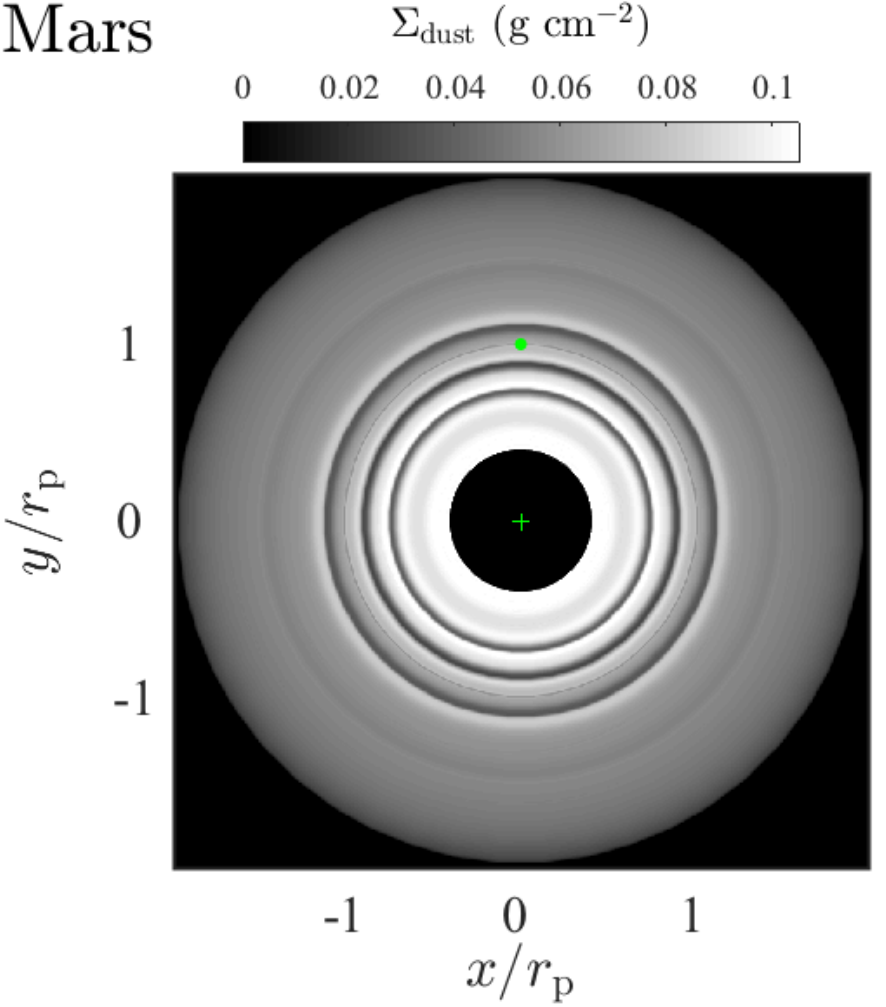}
\end{center}
\figcaption{The dust surface density map for Model Mars at 20,000 orbits. A Mars mass planet ($0.1\me = 3\times10^{-7}\msun$; green dot) can significantly perturb the dust distribution. See \S\ref{sec:implication} for details.
\label{fig:mars}}
\end{figure}

An emerging trend from recent ALMA high resolution disk surveys is that multi-gap structures at 10--100 AU are common in disks around a variety of host stars (S. Andrews, F. Long, private comm.). This demands a robust gap opening mechanism insensitive to disk and host star properties. We argue that such structures can be produced by one or several planets ranging in mass from 0.1 to 10s of Earth masses, located at 1--100 AU. Figure \ref{fig:mars} shows a ``Model Mars'' where a 0.1$\me$ ($3\times10^{-7}\msun$) planet is seen to generate multiple dust gaps over $10^5$ orbits. The low $h/r=0.02$ of this model may characterize disk regions within a few AU from the star, where terrestrial planets in the Solar System reside. Mars-mass planets are capable of being formed by the streaming instability + pebble accretion
\citep{johansen07youdin, ormel17, johansen17, lin18}.

If most multi-gap structures at 10--100 AU are produced by planet-disk interactions, the ubiquity of disk structures implies a ubiquity of planets analogous
to the ice giants (Uranus and Neptune) in our Solar System, and a short time for the formation of their cores (e.g., HL Tau and GY 91 are believed to be $\lesssim$1 Myr old; \citealt{brogan15, sheehan18}). Microlensing surveys
might be able to uncover such a population 
of ``cold Neptunes'' \citep[e.g.,][]{poleski14}. 
Analyzing microlensing survey data available at the time, \citet[Fig. 15; see also \citealt{zhu17ulensing}]{suzuki16} concluded that cold Neptunes are common, with an order unity occurrence rate around main-sequence stars.


\section{Summary}\label{sec:summary}

We carry out two-dimensional, two-fluid hydrodynamical simulations 
of protoplanetary disks with low viscosities ($\alpha\lesssim10^{-4}$)
to study the spacings, depths, and total number of dust gaps opened by one sub-thermal mass planet ($\mplanet<\mth=M_\star(h/r)^3$). Our results provide basic guidelines for interpreting multi-gap structures seen in mm continuum emission. Our main findings are:

\begin{enumerate}

\item Among the observable properties of gaps and rings, gap location is the least affected by observing conditions, finite optical depth, and time evolution (\S\ref{sec:comments}; \S\ref{sec:time}; Figure~\ref{fig:time}). 

\item Gap spacings increase with increasing $h/r$ and decreasing $\mplanet$ (\S\ref{sec:spacing}; Figure~\ref{fig:rp}). We provide empirical fitting functions for the double gap OG1+IG1 (defined in Figure~\ref{fig:example}; Eqn.~\ref{eq:fit:og1ig1}) and IG2 (Eqn.~\ref{eq:fit:ig1ig2}; Figure~\ref{fig:spacing}). 

\item Gaps are opened faster by planets with higher masses,
and in disks with higher $h/r$ (\S\ref{sec:depth}). 

\item More gaps are opened within
a given radius range in disks with lower $h/r$. The number of gaps
is less sensitive to $\mplanet$.

\item The spacings of the double gap in HL Tau \citep{brogan15} and TW Hya \citep{andrews16}, and all three gaps in HD 163296 \citep{isella16hd163296} 
match those of modelled gaps produced by a sub-Saturn mass planet
(\S\ref{sec:applications}; Figures~\ref{fig:realdisks} and \ref{fig:sidebyside}). 

\item A low midplane viscosity ($\alpha\lesssim10^{-4}$) is needed for a sub-thermal mass planet to open multiple gaps (Figure~\ref{fig:alpha}). The Rossby wave instability may develop at the edges of gaps, producing multiple dusty vortices (Figure~\ref{fig:rwi}). A planet as low mass as Mars may significantly perturb the dust disk in terrestrial planet forming regions (Figure~\ref{fig:mars}).

\end{enumerate}


\section*{Acknowledgments}
We are grateful to an anonymous referee for constructive suggestions that improved our paper. We thank David Bennett, Jeffrey Fung, Frederic Masset, Roman Rafikov, Satoshi Okuzumi, and Wei Zhu for discussions, and Eiji Akiyama, Sean Andrews, Crystal Brogan, Giovanni Dipierro, Davide Fedele, Jane Huang, Andrea Isella, and Ryan Loomis for making the ALMA images available. This research used the SAVIO computational cluster at UC Berkeley, as facilitated by Paul Kalas. H.L. and S.L. gratefully acknowledge support by the NASA/ATP and LANL/Center for Space and Earth Science programs.


\appendix
\section{Supplementary Tables and Figures}
\renewcommand\thefigure{\thesection.\arabic{figure}}    
\setcounter{figure}{0}    
\renewcommand\thetable{\thesection.\arabic{table}}    
\setcounter{table}{0}    

\begin{figure}
\begin{center}
\includegraphics[trim=0 0 0 0, clip,width=\textwidth,angle=0]{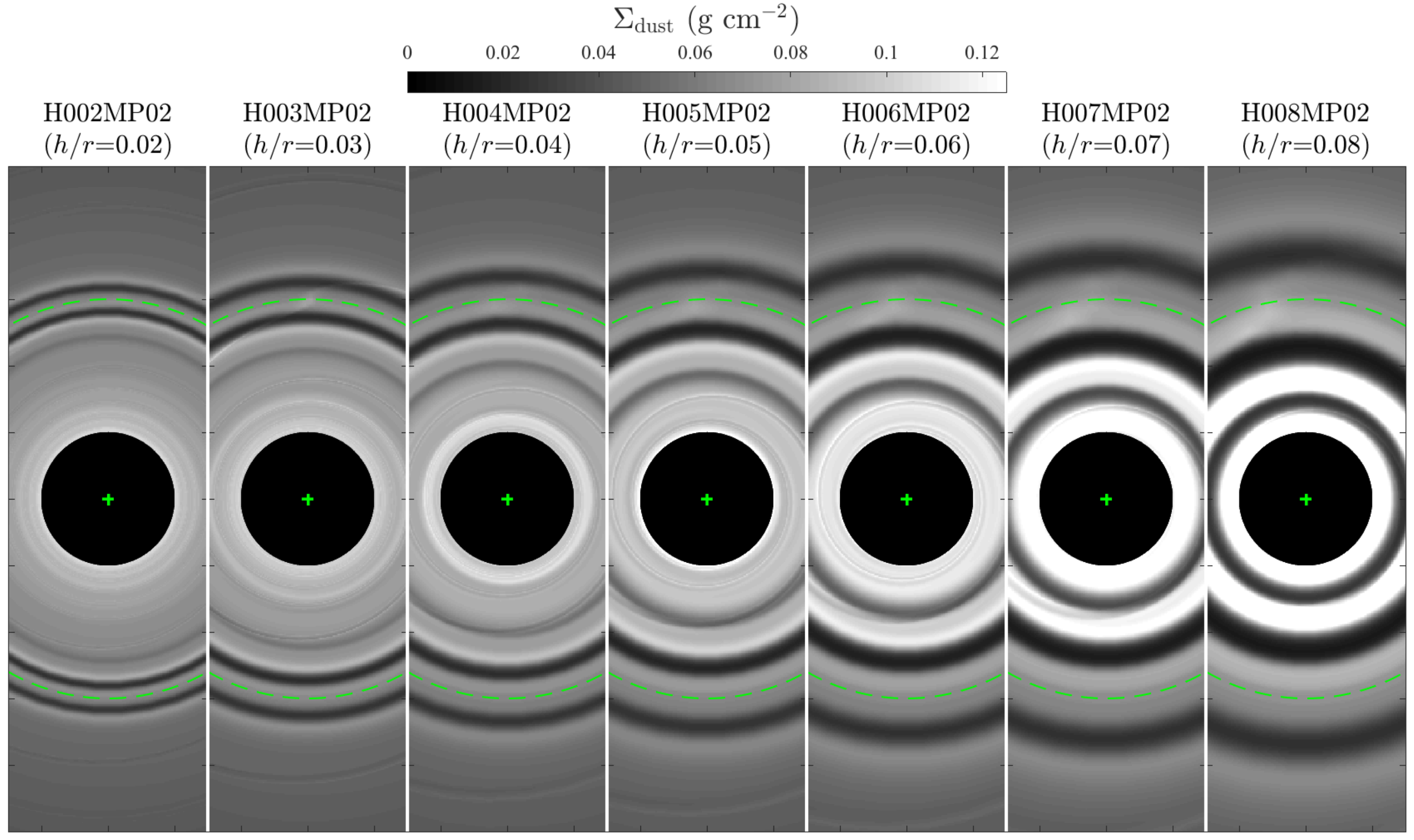}
\includegraphics[trim=0 0 0 0, clip,width=0.729\textwidth,angle=0]{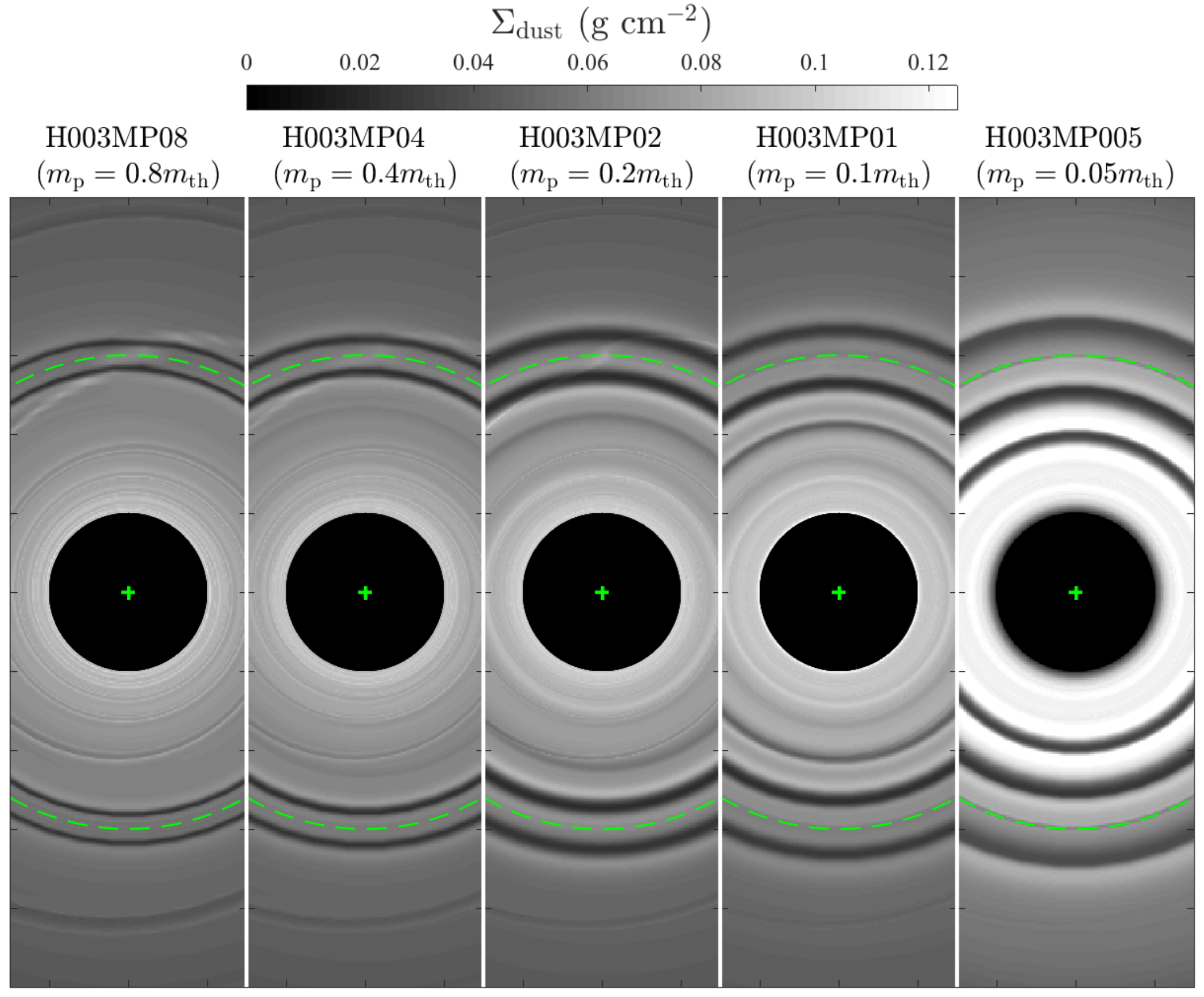}
\end{center}
\figcaption{2D dust maps for two series of models, showing the evolution of dust gap morphologies with varying $h/r$ ({\it top}) and $\mplanet$ ({\it bottom}). See \S\ref{sec:spacing} for details.
\label{fig:mphoverr}}
\end{figure}

\begin{table}[]
\centering
\begin{tabular}{@{}ccccccccc@{}}
\toprule
Model & $r_{\rm OR1}$ & $r_{\rm OG1}$ & $r_{\rm IG1}$ & $r_{\rm IR1}$ & $r_{\rm IG2}$ & $r_{\rm IR2}$ & $r_{\rm IG3}$ & $r_{\rm IR3}$ \\ \midrule
H002MP02 & 1.103 & 1.059 & 0.938 & 0.896 & 0.813 & 0.748 & 0.639 & 0.594 \\
H003MP02 & 1.151 & 1.086 & 0.911 & 0.848 & 0.742 & 0.646 & 0.557 & 0.495 \\
H004MP02 & 1.201 & 1.116 & 0.880 & 0.803 & 0.677 & 0.569 & 0.457 & 0.413 \\
H005MP02 & 1.255 & 1.145 & 0.851 & 0.761 & 0.623 & 0.519 & \dots & \dots \\
H006MP02 & 1.308 & 1.174 & 0.823 & 0.717 & 0.579 & 0.471 & \dots & \dots \\
H007MP02 & 1.360 & 1.205 & 0.794 & 0.673 & 0.535 & 0.425 & \dots & \dots \\
H008MP02 & 1.416 & 1.232 & 0.763 & 0.623 & 0.498 & 0.373 & \dots & \dots \\\midrule
H003MP08 & 1.096 & 1.058 & 0.940 & 0.907 & 0.784 & 0.726 & \dots & \dots \\
H003MP04 & 1.116 & 1.068 & 0.930 & 0.888 & 0.773 & 0.723 & \dots & \dots \\
H003MP01 & 1.174 & 1.109 & 0.883 & 0.821 & 0.706 & 0.640 & \dots & \dots \\
H003MP005 & 1.201 & 1.138 & 0.852 & 0.706 & 0.661 & 0.591 & \dots & \dots \\ \bottomrule
\end{tabular}
\caption{The locations of gaps and rings in units of $\rp$ for models in Figure~\ref{fig:mphoverr}.}
\label{tab:locations}
\end{table}

\begin{table}[]
\centering
\begin{tabular}{@{}ccccccccc@{}}
\toprule
Model & OR1 & OG1 & IG1 & IR1 & IG2 & IR2 & IG3 & IR3 \\ \midrule
H002MP02 & 1.331 & 0.505 & 0.437 & 1.411 & 0.902 & 1.057 & 1.001 & 1.054 \\
H003MP02 & 1.331 & 0.493 & 0.397 & 1.463 & 0.898 & 1.105 & 1.003 & 1.167 \\
H004MP02 & 1.308 & 0.509 & 0.386 & 1.498 & 0.839 & 1.153 & 1.010 & 1.254 \\
H005MP02 & 1.339 & 0.493 & 0.318 & 1.618 & 0.739 & 1.260 & \dots & \dots \\
H006MP02 & 1.371 & 0.485 & 0.266 & 1.777 & 0.640 & 1.407 & \dots & \dots \\
H007MP02 & 1.403 & 0.481 & 0.234 & 1.920 & 0.501 & 1.562 & \dots & \dots \\
H008MP02 & 1.451 & 0.481 & 0.211 & 2.218 & 0.362 & 1.924 & \dots & \dots \\ \midrule
H003MP08 & 1.150 & 0.509 & 0.465 & 1.196 & 1.004 & 1.012 & \dots & \dots \\
H003MP04 & 1.204 & 0.509 & 0.434 & 1.275 & 0.980 & 1.021 & \dots & \dots \\
H003MP01 & 1.228 & 0.499 & 0.353 & 1.387 & 0.668 & 1.174 & \dots & \dots \\
H003MP005 & 1.435 & 0.492 & 0.244 & 1.903 & 0.275 & 1.703 & \dots & \dots \\ \bottomrule
\end{tabular}
\caption{$\sigmad/\Sigma_{\rm dust,0}$ at the locations of gaps and rings for models in Figure~\ref{fig:mphoverr}.}
\label{tab:contrasts}
\end{table}

\end{document}